\documentclass[a4paper]{scrartcl}

\usepackage[english]{babel}
\usepackage[utf8x]{inputenc}
\usepackage[T1]{fontenc}
\usepackage{lmodern,textcomp}
\usepackage{dirtytalk}

\usepackage[a4paper,top=3cm,bottom=2cm,left=3cm,right=3cm,marginparwidth=1.75cm]{geometry}

\usepackage{amsmath}
\usepackage{graphicx}
\usepackage{booktabs}

\usepackage[colorinlistoftodos]{todonotes}
\usepackage[colorlinks=true, allcolors=black]{hyperref}
\usepackage{natbib}
\usepackage{placeins} %for float barrier

\usepackage{setspace}
\title{Adults in the room?}
\subtitle{The auditor and dividends in small firms: Evidence from a natural experiment.}

 \author{Hakim Lyngstad\aa s\\
 BI Norwegian Business School\\
 and \\
  Johannes Mauritzen\footnote{Corresponding author}\\
  NTNU Business School and\\
  NHH Norwegian School of Economics
  }

\begin{document}
\maketitle

% \keywords{}

\begin{abstract}
\textbf{Abstract:} We examine the effect of auditing on dividends in small private firms. We hypothesize that auditing can constrain dividends by way of promoting accounting conservatism. We use register data on private Norwegian firms and random variation induced by the introduction of a policy allowing small private firms to forgo the use of an auditor to estimate the effect of auditing on dividend payout. Identification is obtained by a regression discontinuity around the arbitrary thresholds for the policy. Propensity score matching is used to create a balanced synthetic control. We consistently find that forgoing auditing led to a significant increase in dividends in small private firms.
\end{abstract}

\newpage{}

\section{Introduction}

The level of dividends is a key decision managers make in small- and medium-sized private firms. Dividends reflect a trade-off between compensating owners and maintaining firm liquidity and capitalization. Dividends are also a key variable of interest to policy makers concerned about how financial and accounting regulation may distort firm behavior. The finance and accounting literatures recognize the importance of dividends both as a primary mechanism for compensating shareholders \citep{baker_catering_2004}, as well as a costly signaling mechanism from firm managers to outside stakeholders about the financial condition of a firm \citep{easterbrook_two_1984,miller_dividend_1985, allen_role_2012}.

Arguably, the long-term interests of a private firm and its stakeholders and the short-term preferences of a manager may not always be aligned, even if the manager has a considerable ownership stake. Since equity shares of private firms can be difficult to sell, a cash-constrained owner wishing to extract liquidity from a firm can do so mainly through a dividend payout \citep{bauwhede_financial_2015}.\footnote{A prominent example from US business history of conflict over dividend payouts in a tightly held private firm comes from Koch Industries \citep{leonard_kochland:_2019}. In the late 1970's, Kock Industries was a medium-sized, privately held firm where equity was held primarily by four brothers. While the firm was highly profitable and all the four brothers were wealthy on paper, the firm paid out only a small dividend, preferring to reinvest cash flow to expand the firm. This led to a conflict among the sibling-owners, ending in the shares of two of the brothers being bought out. The firm is currently the largest privately held firm in the US.} Even in the small firms where the manager is the sole shareholder, conflicts could arise between the manager and outside stakeholders such as tax authorities and vendors. Auditors have the purpose of ensuring the completeness, validity and accuracy of transactions as reflected on firm financial statements. In this role, they can influence how a manager balances the trade-off between dividends and retained earnings.

%Asymmetric information between a firm's managers and outside stakeholders could in turn affect dividend behavior \citep{miller_dividend_1985}.

In this article we hypothesize that auditors of private, small- and medium-sized firms constrain dividends by promoting accounting conservatism. The managers and owners of small firms often do not have formal accounting or financial training. Auditors are expected to maintain independence in relation to a firm's strategy and operations, and by law they are not allowed to perform a bookkeeping role. But they do have a duty to report issues with the firm's going concern, such as liquidity and financial robustness. They can also communicate weaknesses in the firm's accounting and internal control processes \citep{downing_audit_2019}.  In other words, the auditor can play the role of the financial \emph{adult in the room}. By pointing out weaknesses in a firm's accounting reports and explaining conservative accounting principles, the auditor can impose a degree of financial discipline on small firms. All else equal, accounting conservatism has the effect of reducing current earnings \citep{watts_conservatism_2003}, which in turn constrains dividends \citep{smith_financial_1979, smith_investment_1992, leuz_role_1998,louis_agency_2015, bradford_conservative_2017}.

To our knowledge, this mechanism of auditor impact is under-explored in the literature. Anecdotally, auditors we have spoken to see this mechanism as plausible for small firms. Small firms often see auditors as financial authority figures. The literature that does exist also seems to provide support for the hypothesis. \citet{manson_value_2001} study management letters by auditors among unlisted companies and find that auditors often provide advice on appropriate accounting methods and policies. % The reason can be explained by a "quiet life hypothesis" \citep{bertrand_enjoying_2003} . However, this requires the firm to relentlessly pursue business opportunities and innovations. Such efforts may be both time consuming, expensive as well as risky. An easier way is to withdraw more dividends, and live a more “quiet life”.

% A large literature in the accounting and finance fields concerns the effect of accounting quality on firm financing. In this literature, we can find a second mechanism that may point in the opposite direction of the \emph{adults in the room} hypothesis.

%Our research question is then not only whether the auditor's role in promoting conservatism can be detected in the form of constraining dividend payout, but also whether that effect will dominate a potential financial substitution effect of accounting quality, which likely moves in the opposite direction.

%\citet{allen_role_2012} finds that private firms that are more dependent on bank financing tend to restrain their dividend payout. Increased information asymmetry due to un-audited financial reports may reduce the external stakeholders ability to accurately monitor the firm's financial position. In turn, this could lead un-audited firms to increase their dividend.

In this article we make use of Norwegian register data on small private firms and a recent implementation of an audit exemption policy for firms under given size thresholds to measure the effect of auditing on dividends. The policy shock is important since accounting conservatism can be difficult to measure directly \citep{chy_real_2021}, and accounting conservatism may be endogenous to the process of determining dividends.

The identifying assumption in our article is that the threshold value for the exemption can be considered arbitrary within a narrow range of firm sizes. Whether a firm is just under or over the threshold can then be considered random. The introduction of this policy can be seen as a natural experiment, with a treatment group (i.e., those not being required to obtain an audit) being randomly assigned. This forms the heart of our identification: Comparing firms just below and above the thresholds to estimate an effect of auditing on dividends. We relax the assumption of random assignment by controlling for indicators of firm size and risk as well as by creating a balanced synthetic control with propensity score matching.

With the inclusion of several variables for firm characteristics, the methodology becomes a regression discontinuity. The estimate of the effect of auditing is obtained by measuring the jump, or discontinuity, at the policy threshold in relation to the overall trend of the size variables. We obtain estimates from specifications with both linear and non-parametric smoothed functions of the size variables. As a robustness check, we estimate difference-in-difference parameters that control for unobserved time-invariant variables.

We find a statistically and economically positive effect of audit exemption on dividends. A firm with operating revenues at the threshold of the exemption policy is estimated to increase their dividend by on average 35,000 - 85,000 NOK (or approximately €3,700-9,000). These estimated average magnitudes are significantly higher than the typical cost of auditing a small firm in Norway. Auditing costs for small firms can vary widely, though a report for the Norwegian Ministry of Finance from 2008 \footnote{\url{https://www.regjeringen.no/no/dokumenter/nou-2008-12/id520230/sec1}} estimates an average cost of around 15,000-20,000 NOK. A report by \citet{langli_evaluering_2015} estimates average cost savings of approximately 14,000 NOK. Thus, the estimates likely reflect a real change in firm behavior rather than just a shifting of saved auditor costs. To put these magnitudes in perspective, the average firm in our sample had operating revenue of approximately 6.7 million NOK. Relative to revenues the average estimated change in dividends is modest, representing only between .5\% and 1.2\% of the operating revenues of a business. However, the mean dividend in our sample was approximately 136,000 NOK. The estimated change in dividends as a percentage of average dividends is substantial.

%We fail to find evidence for alternative explanations, such as firms acting strategically in response to the policy thresholds.

%This study contributes to the literature on auditing quality and corporate governance. As \citet{michaely_corporate_2012} argue, despite broad empirical evidence on dividend payout policies, little is understood about the dynamics of such policies. The results from our analysis provide evidence that an auditor can impose a degree of financial discipline on small firms through improved reporting quality. In turn, an audit exemption may spur some managers to shift resources from investments in the firm to increased pay-outs to the owners. This has implications for both policymakers responsible for auditing regulation, as well as for market actors who wish to evaluate the creditworthiness of firms with un-audited financial statements.

Our contribution is threefold. First, we contribute to the policy discussion of audit-exemption policies. We find that the absence of an auditor leads to an economically meaningful change in firm behavior in the form of increased dividends in small private firms. Second, we add to the literature on accounting conservatism by providing evidence that auditing has the effect of promoting conservative accounting practices and imposing a degree of financial discipline on small private firms that may lack financial sophistication. Finally, we apply the modern theory and methodology of causal modelling with the \citet{rubin_estimating_1974} potential outcomes approach to a current topic within the accounting and auditing literature. The use of such methodologies within accounting research has become increasingly prominent, and this article provides another case study.

In the next section we discuss the institutional setting of our study, including a review of the theoretical mechanisms connecting auditing with dividends in private firms. In section 3, we describe the data and provide descriptive statistics. Section 4 provides a discussion of the assumptions necessary for a causal interpretation of our results and related challenges. Section 5 describes the regression discontinuity design and presents results. Section 6 extends the regression discontinuity model to include a matched control group that provides improved balance. Section 7 discusses results from a difference-in-difference estimation as a robustness check. Section 8 concludes with a summary of the results and implications.

\section{Institutional Setting}

While Norway is not an EU member, it is a member of the European Economic Area. Norway is bound to most European accounting and auditing laws and standards established by 4th and 7th European commission directives (\say{the accounting directives}) from 1978 and 1983. Norway is also a small open economy, where external trade makes up a large portion of GDP and where many firms have foreign customers, suppliers, investors, and owners. Norwegian auditing standards are largely in line with those from the International Standards of Auditing (ISA) \citep{downing_audit_2019}.

Norway has some institutional features that differentiates it from other countries. Much of the literature on auditing in private firms uses data from Anglo-Saxon countries, with a tradition for common-law judicial systems and relatively high litigation risk. Norway has a civil judicial system more in line with much of continental Europe, with lower litigation risk. This could in theory affect auditor independence, as the threat of litigation is sometimes seen as a check on auditors breaching their obligation of independence. But a study by \citet{hope_auditor_2010} failed to find evidence for breaches of auditor independence in Norway.

%The rationale behind audit exemption rules is that the external costs of audit exemption for small firm are  minimal, and that small firms may have private information about whether the private benefits of an audit outweigh the costs.

The audit exemption rule was implemented in Norway as a part of a larger policy of reducing the bureaucratic and administrative costs to firms, especially small firms and start-ups. This policy goal was codified into law, requiring the government to reduce the bureaucratic costs of conducting business by at least 3 billion NOK per year. The audit exemption was the largest component of this policy, with an estimated cost reduction of 2 billion NOK per year \citep{langli_evaluering_2015}.

Most EU countries have long traditions for audit exemptions. Figure \ref{table:thresholds} shows thresholds for the Nordic countries as well as the UK. Small countries tend to have correspondingly small threshold values compared to the UK. Norway's thresholds were set to a maximum balance sheet of approximately €2 million (20 million NOK) and a net turnover of approximately €500,000 (5 million NOK), and no more than 10 employees.\footnote{From 2018, these were increased to 6 and 23 million NOK, though we only have data through 2015.} In some countries the audit exemption thresholds coincide with thresholds associated with other policies. To our knowledge, the thresholds in Norway are only used for purposes of determining whether a firm qualifies for an audit exemption.

\begin{table}[h]
\label{tbl:int_comparison}
\caption{\label{table:thresholds} Audit exemption thresholds in  Norway, Nordic EU Member States, and the UK. Obtained from the Federation of European Accountants.}
\begin{center}
\begin{tabular}{lccc}
\toprule
Country &Total assets (€) &	operating revenue (€) &	Number of employees\\
\midrule
Norway&		2,000,000&	500,000 &	10 \\
Sweden &	150,000 &	300,000&	3	\\
Denmark&	4,837,000&	9,674,000&	50	\\
Finland&	100,000&	200,000&	3	\\
Iceland&	1,400,000&	2,800,000&	50	\\
UK&		6,541,000&	13,082,000 &	50 \\
\bottomrule
\end{tabular}
\end{center}
\end{table}

The threshold conditions for the Norwegian policy are cumulative, meaning that all conditions must be met before a firm can forgo auditing.  To issue the proxy, a two-thirds majority of the votes and share of the capital represented at the annual general  meeting  is  required.\footnote{But, both  the  Norwegian  law  and  the  EU  Fourth  Company  Law  Directive  specify  exemptions  from voluntary  audit regulations.  For  instance,  parent companies are required to submit audited financial statements regardless of firm size. As a consequence, a majority of subsidiary companies are audited, regardless of the flexibility the legislation  provides. Some industries are also entirely exempted from this legislation (banks,  insurance  companies,  law  firms, auditors,  providers  of  financial  services  and  other  entities  under  scrutiny  of financial  regulators.} Status in any given year is determined from the previous year's financial results. The thresholds are ``hard'', meaning that they are not adjusted for firms that move across them from one year to another.

Norway is an interesting case study because it is a late adopter of the audit exemption policy. The UK, for example, already introduced an audit exemption in 1994 \citep{collis_demand_2004}. By contrast, the Norwegian policy was rolled out first in 2011. Audits can be a significant expense for small- and medium-sized firms. \citet{kausar_real_2016} estimate that an audit can cost 6\% or more among private small- and medium-sized European firms. Because Norway has had an auditing requirement for all firms up until 2011, a competitive market for audit services has developed with transparent pricing. Thus, we can compare our estimated magnitudes to reasonable estimates of the cost of obtaining an audit.

\subsection{Theory of auditing and dividends}

The hypothesis that auditing can reduce dividends by promoting accounting conservatism is, to our knowledge, new to the literature. Yet it can be seen as a combination of component ideas that have strong backing in both theoretical and empirical research. First, there is the question of whether mandatory auditing has any meaningful effect on firm  behavior at a broad level. Furthermore, we are concerned with evidence that auditing promotes conservative accounting practices in firms. Finally, our proposed mechanism involves accounting conservatism reducing dividends by way of constraining current earnings.

An extensive literature finds evidence that auditing has real financial effects on firms \citep{blackwell_value_1998, allee_demand_2009, kim_voluntary_2011, minnis_value_2011}. \citet{kausar_real_2016} is methodologically similar to our own study, making use of the implementation of the UK audit exemption policy as a natural experiment. They find that firms obtaining voluntary audits improved their investment and financial performance. \citet{lennox_voluntary_2011} also study the effects of the UK audit exemption policy and finds that firms that voluntarily submitted to an audit received a higher credit rating compared to when there was a mandatory audit. In contrast to our own study, both these articles have a theoretical focus on the signaling effect of voluntary audits.

Several studies find evidence that auditing imposes financial discipline and promotes conservative accounting practices in firms. \citet{clatworthy_impact_2013} and \citet{downing_audit_2019} find that audited firms had higher compliance with tax and auditing regulations. \citet{krishnan_audit_2003} find a positive relationship between audit quality (as proxied by a big-six audit) and \say{opportunistic} accruals by management. \citet{krishnan_audit_2005} and \citet{zhang_audit_2007} find positive associations between measures of auditor quality and fewer internal control problems. \citet{chy_real_2021} find evidence that auditor conservatism reduced investments in innovation. \citet{badolato_audit_2014} find that higher relative status of an auditor reduces earnings management. This is particularly relevant in our case as the firms in our sample are small and more likely to be financially unsophisticated. Presumably the auditor will have a high relative status in many such situations.

The second link in our hypothesis is between accounting conservatism and dividends. The accounting literature has long acknowledged that a conservative determination of net earnings restricts dividends in practice \citep{smith_financial_1979, smith_investment_1992}. \citet{leuz_role_1998} argues that this relationship can be explained by the implicit constraints the accruals process puts on investment decisions made to generate a certain timing of future cash flow. Subsequent studies have supported the role of accounting conservatism in constraining net earnings and in turn dividends \citep{watts_conservatism_2003, andres_dividend_2009, louis_agency_2015, bradford_conservative_2017}.

The hypothesis that forgoing auditing will on average lead to higher dividends among small firms is not ex-ante obvious. The theory on audit quality suggests that the absence of auditing could lead to increased retained earnings and lower dividends. If the absence of an auditor reduces the quality of the accounts, this could negatively affect financing by increasing the information asymmetry between a firm and its lenders. Empirical studies have shown that higher auditing quality leads to both lower cost of debt \citep{francis_market_2005,bauwhede_financial_2015} and equity \citep{hribar_effect_2004}. As the cost of capital decreases, the firm may pay out higher dividends compared to firms choosing to forgo auditing. Several studies provide empirical support for this hypothesis \citep{koo_effect_2017, lawson_earnings_2016, caskey_dividend_2013}.

\section{Data}

Our data is from the Norwegian Centre for Corporate Governance research database, which consists of all Norwegian firms for the sample period (2010-2015). We limit the sample to firms that were active during the entire period 2005-2015. This limits the sample to 27,126 unique firms. We remove micro firms, with revenues or total balance of less than NOK 500,000 (approximately €50,000). We do not filter based on the number of employees since after filtering on revenues and assets, no firms had large deviations from the threshold on number of employees. We removed firms that do not report revenues, inventory, account receivables, or accounts payable each year in the sample period. Some industry groups are not eligible for an audit exemption and have not been included. This includes public limited corporations, financial firms, law firms, foundations, lottery firms, drugstores, and accounting and auditing firms. After these exclusions, we are left with 5,293 firms. We also remove firms with revenues of over NOK 15 million and a total balance of over NOK 50 million. We are left with 3,025 unique firms. Finally, the remaining firms are checked against the official Norwegian firm registry ("Brønnøysundsregistrene"), leaving 2,866 firms.

We loosen the assumption of random treatment assignment by controlling for a host of variables. Firm size is important as it signals that; i) managerial ownership normally decreases with increased firm size and ii) there may be fixed costs associated with starting an audit relationship, which will be relatively smaller for larger firms \citep{chow_demand_1982,fama_disappearing_2001,sharma_independent_2011}. We measure firm size as both total assets and operating revenue. These variables also determine eligibility for the audit exemption. By controlling for firm size in both the treatment and control groups, we measure the discontinuity at the threshold rather than the average difference between groups. Given the importance of controlling for firm size between treatment and control groups, we present results with both linear controls and non-parametric smoothed controls for firm size.

Agency theory suggests that high-risk firms are less likely to pay dividends, as they tend to have reduced access to capital markets and need to be self-financed to a higher degree \citep{ho_dividend_2003}. We measure firm risk as the standard deviation of the return on assets (ROA), where ROA is defined as earnings before interest and taxes plus interests, divided by total assets.

According to the \citet{jensen_agency_1986} free cash flow hypothesis, higher cash holdings should be related to higher dividend payouts. Cash flow is measured as the sum of operating income and depreciation. We include firm mean cash flow across the years 2011-2015 in our analysis. Since cash flow uncertainty may constrain dividends, the standard deviation of cash flow is also included.

The profitability of a firm is important in determining the dividend. However we consider the reported income to be a post-treatment variable and co-determined with dividends, an issue we discuss below. But we do include lagged operating income as a control variable.

Leverage may also influence the propensity to pay dividends, as firms must ensure that they have sufficient liquidity to meet short- and long-term obligations. We measure leverage as total debt divided by total assets.

Finally, we control for sector and year fixed effects in the analysis. Table \ref{tbl:full_specification} shows the names and descriptions of the variables included in the regressions.

\begin{table}[!htbp] \centering
\caption{The table presents all variables and their definitions included in the regression models.}
\label{tbl:full_specification}
\begin{tabular}{ll}
\\[-1.8ex]\hline
\hline \\[-1.8ex]
Variable & Definition\\
\hline \\[-1.8ex]
dividend & Dividends per firm per year, normalized$^*$\\
noAudit & Indicator of whether a firm chose to forgo an audit\\

operating revenue  & Operating revenue per firm per year, \\
        & in thousands of Norwegian kroners, normalized \\

total assets  & Total assets per firm per year, \\
        & in thousands of Norwegian  kroners, normalized \\

operating income (t-1) & 1-year lagged operating income per firm per year, \\
            & in thousands of Norwegian kroners, normalized\\

leverage  &  Ratio of total debt to total assets,\\
      &    per firm per year, normalized \\

risk (roa sd) & Risk as proxied by within-firm standard deviation \\
              & of return on assets \\
              & (operating income divided by total assets), normalized \\

cash flow (mean) & Within-firm mean of cash flow, normalized\\

cash flow (sd) & Within-firm standard deviation of cash flow, \\
                & normalized\\
\hline \\[-1.8ex]
\multicolumn{2}{l}{\textit{$^*$Normalized: Subtracting the mean and dividing by the}}\\
\multicolumn{2}{l}{\textit{standard deviation of the variable.}}
\end{tabular}

\end{table}

Table \ref{tbl: descriptive_stats} presents summary statistics for the variables used in the analysis. Statistics are calculated from firm-year observations from 2011 through 2015, which are the years in our sample where the audit exemption was instituted. Many firms pay out 0 in dividends at least some years. By design, all firms have significant amounts of revenue, however some firms have years with negative operating income. There is a wide range for both leverage and return on assets, but the 25th and 75th percentile statistics show that most firms lie within a much narrower range. Finally, the proportion of firms in a year that are eligible and choose to forgo an audit is 14\% of the total.

 \begin{table}[!htbp] \centering
 \caption{Summary statistics of 14,330 firm-year observations from 2011-2015.}
\label{tbl: descriptive_stats}
\begin{tabular}{lcccccc}
\\[-1.8ex]\hline
\hline \\[-1.8ex]
Statistic & \multicolumn{1}{c}{Mean} & \multicolumn{1}{c}{St. Dev.} & \multicolumn{1}{c}{Min} & \multicolumn{1}{c}{Pctl(25)} & \multicolumn{1}{c}{Pctl(75)} & \multicolumn{1}{c}{Max} \\
\hline \\[-1.8ex]
Dividend (1k NOK) & 141 & 390 & 0.00 & 0.00 & 50 & 10000 \\
Operating Revenue (1k NOK)  & 6667 & 3319 & 503 & 3948 & 9098 & 25742 \\
Total Assets (1k NOK)  & 4028 & 3379 & 500 & 1918 & 4997 & 42977 \\
Leverage  & 0.60 & 0.44 & -0.15 & 0.34 & 0.79 & 20.65 \\
Operating income (lagged) & 251 & 581 & -6360 & -10 & 431 & 8219 \\
Return on Assets & 0.09 & 0.18 & -5.44 & 0.01 & 0.17 & 2.0 \\
noAudit & 0.14 & 0.34 & 0.00 & 0.00 & 0.00 & 1.0 \\
\hline \\[-1.8ex]
\end{tabular}

\end{table}

\begin{figure}
  \caption{Correlations among the variables of interest. A star indicates a correlation coefficient that is statistically significant at the 95\% significance level.}
 \centering
 \includegraphics[width=1\textwidth]{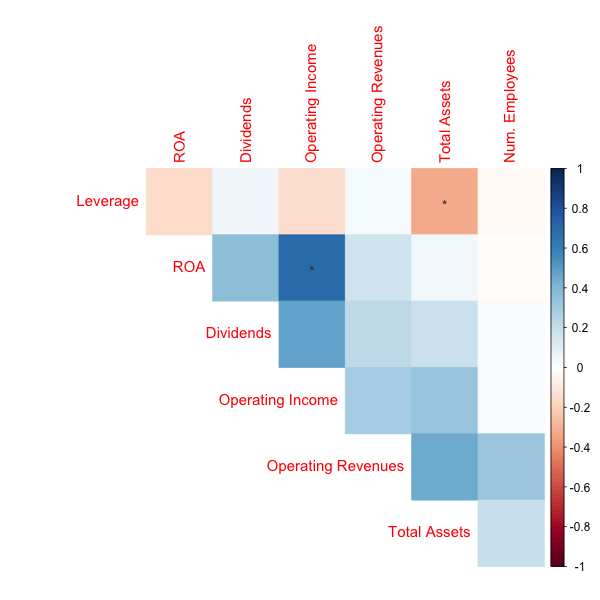}
 \label{fig:correlationPlot}
 \end{figure}

Figure \ref{fig:correlationPlot} shows correlations among the variables of interest. A star within a square indicates a correlation coefficient that is statistically significant at the 95\% confidence level. Notably, the correlation coefficient between operating income and dividends is relatively high at approximately .50, though this is not statistically significant.  The correlation between dividends and operating revenues, total assets, and number of employees, which are used as thresholds for the auditor exception rule, are substantially lower.

\begin{figure}
\centering
\caption{The figure shows scatter plots for each year of firm operating revenue versus total assets. The red lines represent the thresholds for eligibility to forgo auditing. The orange dots are firms that have forgone auditing. A limited trial run began in 2011, with a full roll-out in 2012. Operating revenue is nearly always the binding constraint for eligibility.}
\includegraphics[width=1\textwidth]{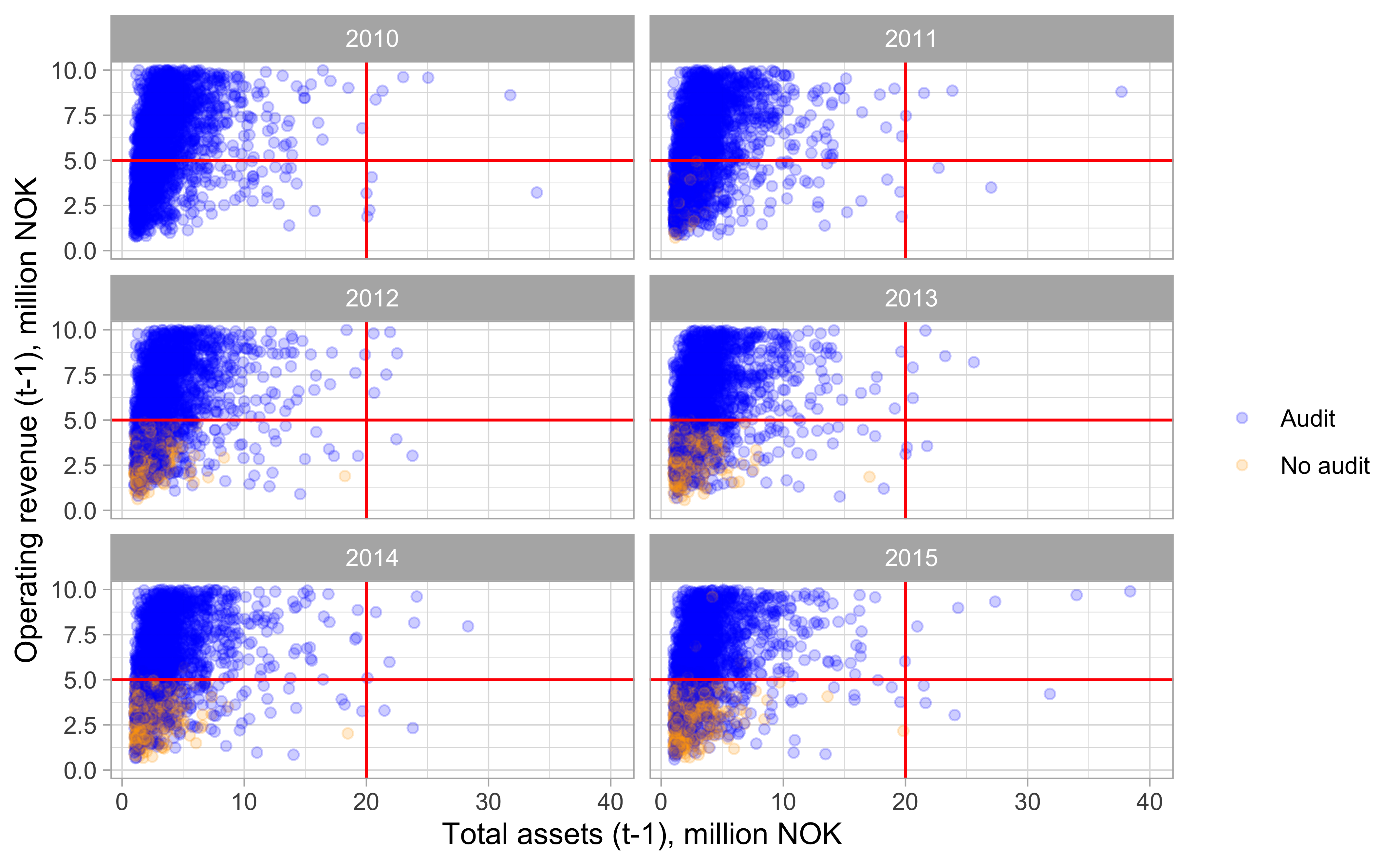}
\label{fig:scatter_size}
\end{figure}

Eligibility for the exemption policy is established by way of three thresholds: Operating revenue, total assets, and number of employees. Figure \ref{fig:scatter_size} shows a scatter plot of all the firms in our data by operating revenue and total assets. The red lines represent the policy thresholds for firm size. Orange dots represent firms that chose to forgo auditing in that year. The policy had not been implemented yet in 2010, and only a limited roll-out happened in 2011. 2012 was the first year with a full roll-out, where all firms under the thresholds could forgo an audit. The figure makes clear that in most cases the binding constraint for eligibility is operating revenue. The number of firms who come under the operating revenue threshold but over the total assets threshold is as little as two firms (2012-2014) and at most four firms (2015). Thus, the effect of total assets as a binding constraint is negligible.\footnote{The threshold for number of employees is more problematic. Figure \ref{fig:income_employee_scat} in the appendix shows a scatter plot of number of employees and operating revenue. Again, operating revenue is the binding constraint for the large majority of firms. But we also see that many firms that come under the operating revenue threshold but over 10 employees nonetheless have been able to choose not to get an audit. This is because the threshold is for 10 full-time equivalent employees, while the data in the sample is per head. Thus two 50\% employees will count as 1 employee towards the maximum, but will appear as two employees in the data. Not surprisingly, many small firms in the sample rely on part-time employees. This makes using the number-of-employees threshold impossible in the regression discontinuity design. This is likely only a minor problem since only a small number of firms that come under the two other thresholds will be close to or above the 10 full-time-employee threshold.}

A concern is that firms are intentionally adjusting their operating revenue to come just below the threshold. A recent literature finds evidence for such strategic manipulation of reported firm size to avoid disclosure or auditing \citet{bernard_size_2018}. In our case, strategic size management may be particularly concerning since manipulating accounting data to come under the thresholds is contrary to conservative accounting practices and might introduce a bias in our estimation.

\begin{figure}
\centering
\caption{The figure shows histograms of operating revenue before and after the implementation of the auditing exemption policy. The vertical black line represents the threshold. The scale is limited to be close to the threshold to highlight any potential bunching. In the case of firms intentionally adjusting their operating revenue to come just under the threshold, we would expect to see "bunching" of firms on the left side of the threshold. No pattern of bunching is apparent.}
\includegraphics[width=1\textwidth]{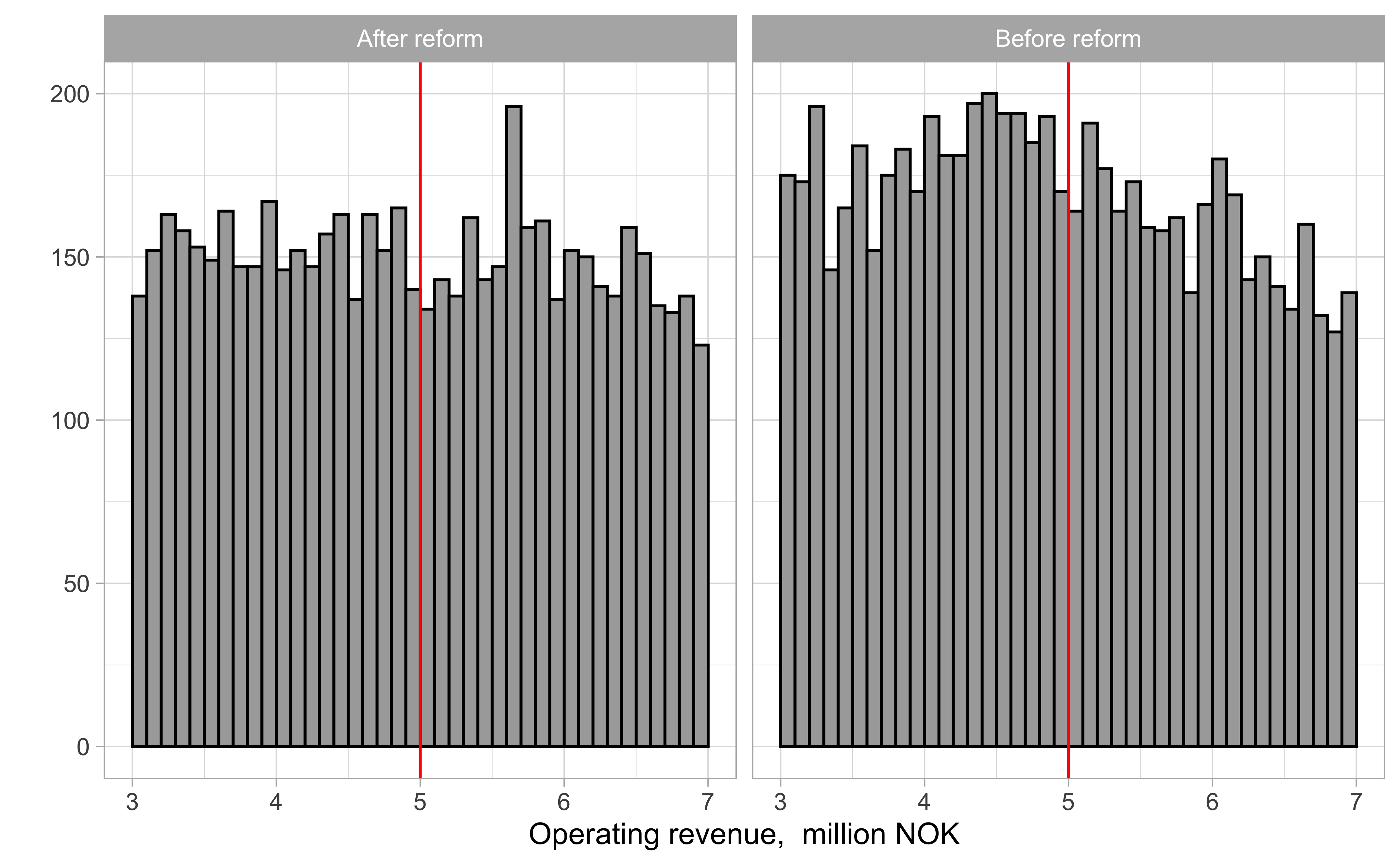}
\label{fig:bunching}
\end{figure}

Identifying size management is straight forward. If such manipulation exists, then it should have the effect of creating a discontinuity at the threshold, as those just over the threshold have an incentive to manage their accounts downward. We focus on the operating revenue threshold, as it is binding for most observations. Figure \ref{fig:bunching} shows histograms of operating revenue before and after the implementation of the policy, with vertical black lines representing the threshold values. The limits of the x-axis are constrained to values close to the threshold. If firms were acting strategically in response to the exemption policy, we would expect to see ``bunching'' of frequencies just below the threshold after the implementation of the policy. No pattern of bunching is apparent.

\FloatBarrier
\section{Identifying the effect of an audit exception}

We use a regression discontinuity approach to identify the effect of an audit exemption on dividends. The identifying assumption is that the threshold set for defining what firms are small enough to forgo an auditor is arbitrary within a narrow range of firm size.

We can first consider a theoretical scenario where an experimental randomization can be performed. In the case of random assignment to a control and treatment groups, unobserved variables that might otherwise bias the results satisfy the criteria of ignorability \citep{rubin_estimating_1974}. Using the notation of \citet{imbens_causal_2015}, we let $Y_{i1}$ represents the potential outcome if the firm, i, is not audited, and $Y_{i0}$ represents the potential outcome if firm i is in the control group, which is audited. Treatment, in this case, is the introduction of the auditing exemption, represented by $T_i=1$. Given randomization, the average treatment effect can be estimated as in equation \ref{eq:simple_treatment}. In other words, the claim is that the randomization gives the possibility to make an estimation from a counterfactual: The difference in the dividend pay-out of a given firm $i$ under both a scenario where they are, and are not, audited.

\begin{align}
\tau &= E(Y_{i1|T_i = 1}) - E(Y_{i0|T_i = 0}) \\
     &= E(Y_i|T_i=1) - E(Y_i|T_i=0)
\label{eq:simple_treatment}
\end{align}

A complication for our case is that it is not auditing itself that is randomized, but rather the option to be exempt from auditing. The quantity of interest then becomes the Average Treatment Effect for the Treated (ATT), as in equation \ref{eq:att}. In words, this is the average difference in the dividend payout of a given firm $i$ who is not audited ($Y_{i1})$ with the counterfactual of the same firm that obtains an audit ($Y_{i0}$), given the introduction of an audit exemption ($T_i=1$).

\begin{equation}
\tau|(T=1) = E(Y_{i1}|T_i=1)-E(Y_{i0}|T_i=1)
\label{eq:att}
\end{equation}

We have several reasons for being interested in the ATT versus the pure average treatment effect. The first is that a pure treatment effect is impossible to estimate if firms had a choice of whether to forgo an auditor. From a policy evaluation perspective, it is also the ATT effect that is relevant. The counterfactual scenario of all firms under the threshold forgoing an auditor is not of particular interest in evaluating a policy where firms are in fact given the choice of whether to obtain an auditor.

\begin{figure}
\centering
\caption{Quadrants I and II represent firms that are below the size-thresholds and therefor eligible to forgo auditing. Quadrants III and IV represent firms not eligible to forgo the auditor. To estimate the average treatment on the treated (ATT) effect, we wish to compare firms in quadrant I (treatment) with firms in quadrant IV (control). We observe which firms are eligible to forgo auditing (bottom quadrants), and we observe those who forgo auditing if given the chance (quadrant I). However, for firms not eligible to forgo auditing, we do not observe whether they would forgo auditing if given the possibility (quadrant IV). Instead, we must either control for the characteristics that determine whether firms are in quadrants III and IV or estimate a synthetic control group.}
\label{fig:identif_map}
\includegraphics[width=.8\textwidth]{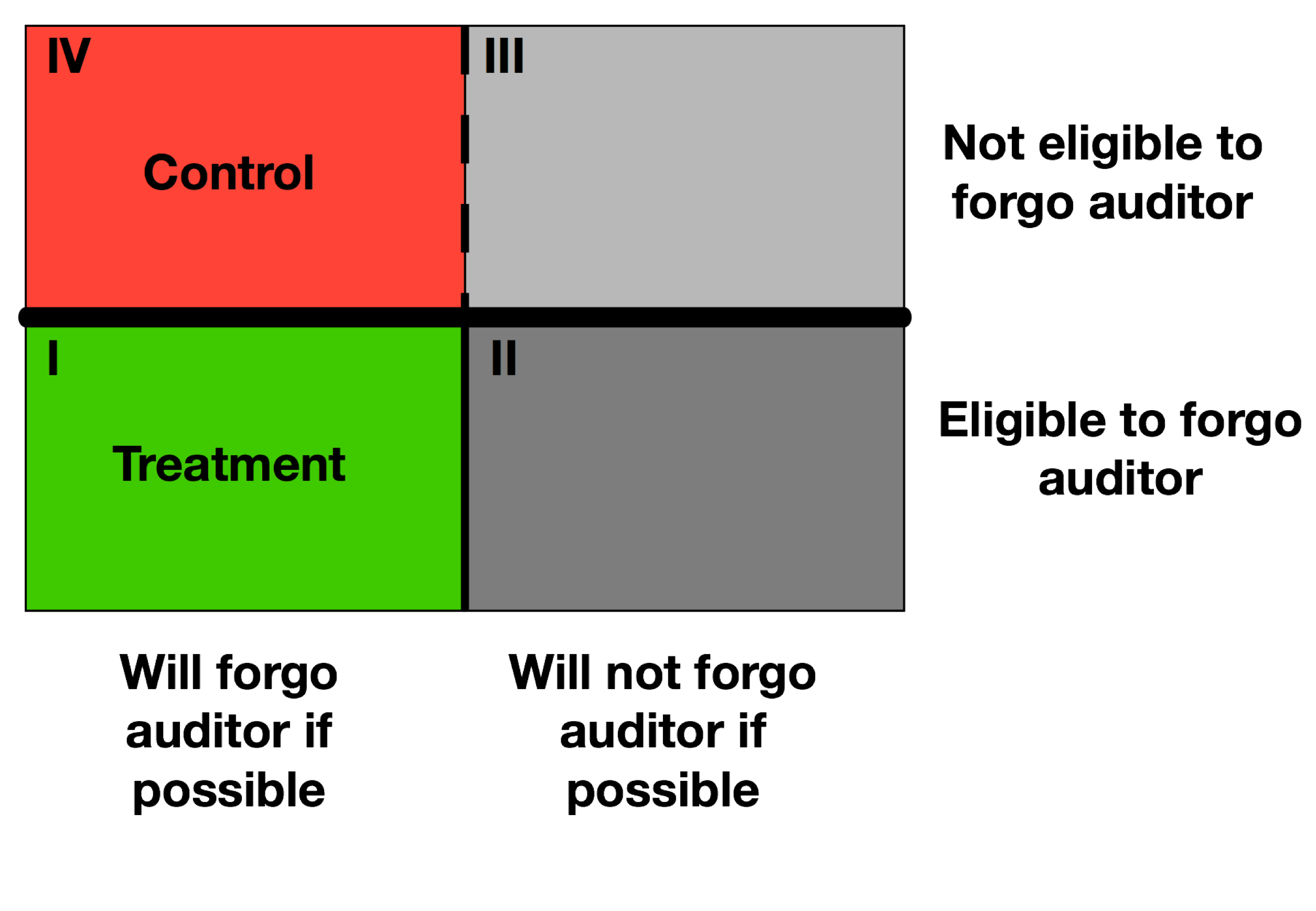}
\end{figure}

A problem with the ATT as written in equation \ref{eq:att} is that the counter-factual $E(Y_{i0}|T_i=1)$--the outcome for the treated observations if they did not receive the treatment--cannot be directly observed. The problem is that we do not observe which firms above the threshold would have forgone an auditor if given the choice (quadrants III and IV in figure \ref{fig:identif_map}). If the entire group above the threshold is used as a control, then this could confound the results. In particular, the estimation may pick up on a signal and screening effect. Since an exemption policy makes choosing to get audited observable, this can be used as a signal to stakeholders about the company's financial situation. In a similar fashion, since audits are costly, choosing to get audited can have a screening effect, with high-quality firms willing to pay the auditing cost to demonstrate their financial robustness independent of the effect of the actual financial reports \citep{kausar_real_2016}.

In attempting to estimate the real financial reporting effects of an audit, we compare those firms that chose not to obtain an audit with a control group that was above the threshold and thus was required to obtain an audit. The firms that voluntarily chose to be audited are excluded. The key empirical challenge in this article is controlling for the factors that affect whether a firm above the threshold would choose to forgo an audit if they could.

Including control variables can help alleviate the potential bias that may be introduced by this self-selection but may not be sufficient to make the sample balanced between the treatment and control. Therefore, we also use propensity score matching to create a synthetic control group with improved balance.

By making use of the full panel of our data and a difference-in-difference identification strategy, we can also control for unobserved variables that may be correlated with treatment assignment. Instead of only comparing above and below the threshold after the introduction of the exemption policy, we compare changes in firms' dividend between the years directly before and after the introduction of the policy and then compare average changes between firms in the treatment and control groups. In this way, unobserved variables that are correlated with the treatment but that are time-invariant are differenced out. The difference-in-difference strategy comes at the cost of reduced efficiency, as the differencing has the effect of discarding information from the data.

\FloatBarrier
\section{Regression discontinuity design}

Changing to standard regression notation, the regression discontinuity design can be written as:

\begin{equation}
DIV_{it}=\tau NoAudit_{it}  + f(\mathbf{SIZE}) +  \mathbf{X \beta} + \mathbf{Z\gamma} + \epsilon_{it}
\label{eqn:reg_disc}
\end{equation}

$DIV_{it}$ represents the dividends of firm $i$ in year $t$. Here the treatment effect is represented by the coefficient $\tau$ on the indicator $NoAudit_{it}$, whether firm i forgoes an audit in year t. $ f(\mathbf{SIZE})$ represents the function of the firm size variables that determine the thresholds for the audit exemption. $\mathbf{X \beta}$ and $\mathbf{Z\gamma}$ represent vectors of observed and unobserved variables and corresponding vectors of coefficients. Examples of unobserved variables that might affect dividends include manager personality and risk appetite.

In this specification, $\tau$ is the discontinuity at the threshold value of the policy, conditional on the control variables. The estimate of $\tau$, $\hat{\tau}$, is only unbiased if the selection into the treatment is not dependent on the unobserved covariates $\mathbf{Z}$: $E(\tau|\mathbf{f(SIZE), X, Z}) = E(\tau|\mathbf{f(SIZE), X})$.

Covariates are chosen so as not to control for post-treatment variables, which will tend to bias the estimated coefficients towards zero \citep{gelman_data_2006}. Controlling for earnings and cash flow as well as other flow variables in the financial accounts could then partially control for the effect on dividends that we wish to estimate. On the other hand, variables like earnings and cash flow also have a straight-forward relationship to dividends as these are determinants of a firm's ability to pay dividends in the first place. We include control variables for firm size: Operating revenue, total assets, and number of employees. Stock variables, like leverage, reflecting a history of the firm's activities are also included. Flow variables, like cash-flow, that fully reflect year-to-year variation are not included. Instead, we include firm average and standard deviation of cash-flow over the years 2005-2015. We also include the one-year lag of operating income, though this variable does not materially alter the estimation of the coefficient on the \emph{noAudit} indicator. Since the variable for within-firm average cash flow already controls for the general level of profitability of a firm (cash flow is defined as operating income plus depreciation), we omit the lagged operating income variable following the first set of discontinuity regressions.

\subsection{Results from the regression discontinuity design \label{reg_discontinuity}}

Table \ref{tbl:discon} shows results from the regression discontinuity estimation. Data is for the period 2011 through 2015, when the policy was in place. In the first column, only an intercept term, four year fixed effects and the treatment indicator are included. The treatment indicator is shown to be negative. This estimation is equivalent to a computation of the average difference in dividends between firms in the sample that were- and were not audited. But this is a biased estimate of the effect of auditing. Dividends and the \emph{noAudit} indicator are likely correlated with omitted variables. In particular, the absence of measures of firm size, which are positively correlated with dividends, and which determine eligibility for the auditing exemption will negatively bias the estimate. Including control variables alleviates this bias. Operating revenue is important since this is the measure of firm size that is the binding determinant of eligibility for an auditing exemption for most firms. With the inclusion of covariates, we can interpret the \emph{noAudit} coefficient as the jump, or discontinuity, at the eligibility threshold for firm size.

In the second column, we control for indicators of firm size, profitability, and risk. The treatment effect is now estimated to be positive and significant at the 10\% level. The estimates in the second column include firms in the control group who were eligible to forgo an audit but chose to obtain an audit (quadrant II in figure \ref{fig:identif_map}). This is a departure from the goal of estimating the average treatment on the treated (ATT) effect.

In the third column we exclude observations who self-selected out of the treatment group. The estimated coefficient on the \emph{noAudit} indicator increases to .15 standard deviations. This corresponds to approximately 60,000 NOK (€6,000). This provides insight into the bias that self-selection induces in the treatment. The direction of the coefficient change indicates that those firms who are under the threshold but still choose to be audited, tend to pay out a lower dividend than firms mandated to obtain an audit. If the same pattern holds for firms firms over the threshold (quadrants III and IV in Figure \ref{fig:identif_map}), then the regression results for the coefficient on the \emph{noAudit} indicator would have an attenuation bias and could be considered a conservative estimate. On the other hand, a direct regression on all firms eligible to forgo an audit finds no significant relationship between dividends and choosing to be audited (see table \ref{tbl:underThreshold} in Appendix B), indicating that that the bias from self-selection is modest.

In the fourth column, 94 sector fixed effects are included in the model. This appears to have little effect on the estimated coefficient on the \emph{noAudit} indicator.

\begin{figure}

\caption{The figures show firm-year observations in the data. The plot on the left shows the full sample of data used. The plot on the right shows only the narrow subset of data close to the operating revenue threshold, which is also represented by the data between the dotted lines on the left plot. The plots do not include firms that voluntarily choose to obtain an audit if they were below the threshold.}
 \label{fig:narrow_plot}

\begin{minipage}{.40\textwidth}
  \centering
  \includegraphics[width=1\linewidth]{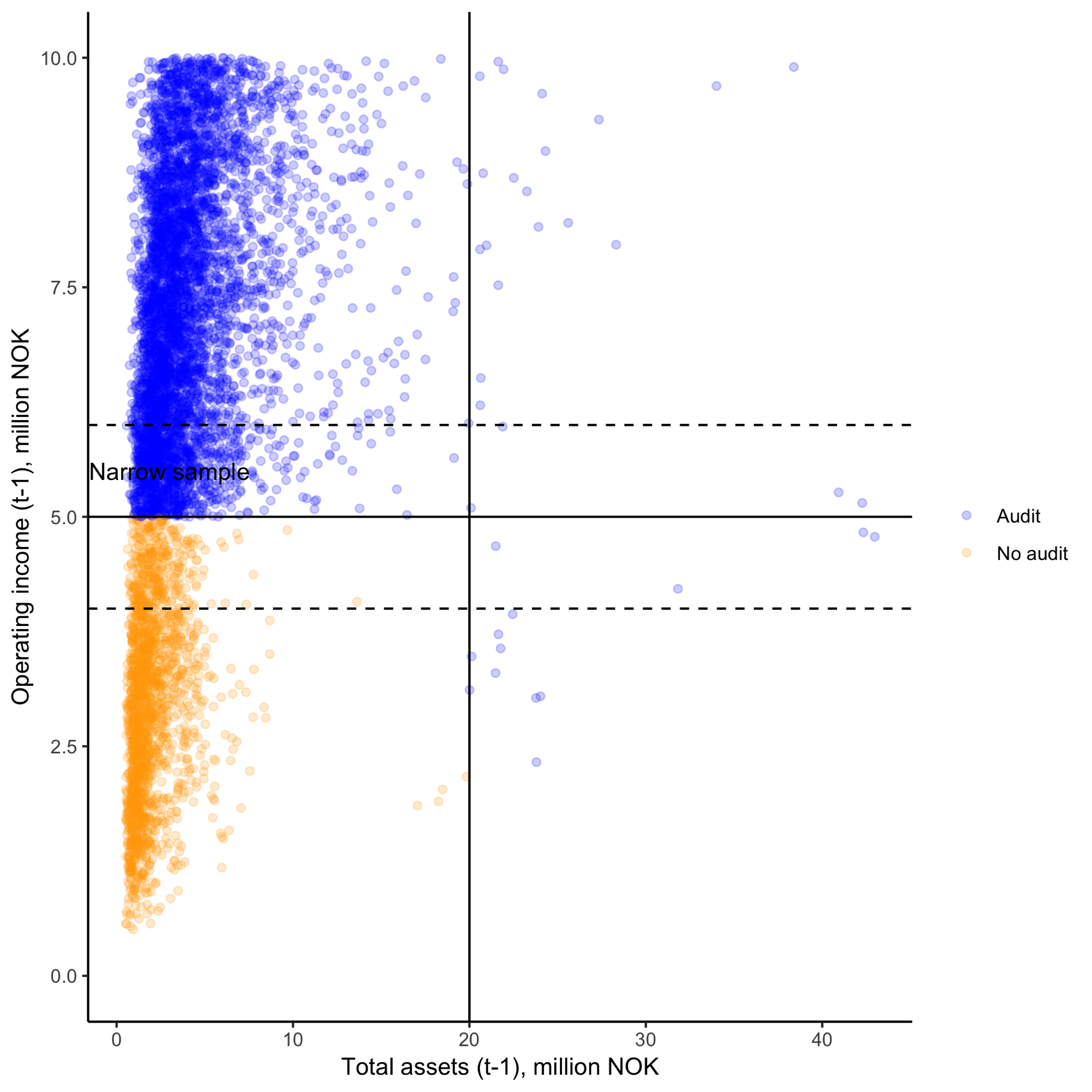}
 \end{minipage}\qquad
\begin{minipage}{.40\textwidth}
  \centering
  \includegraphics[width=1\linewidth]{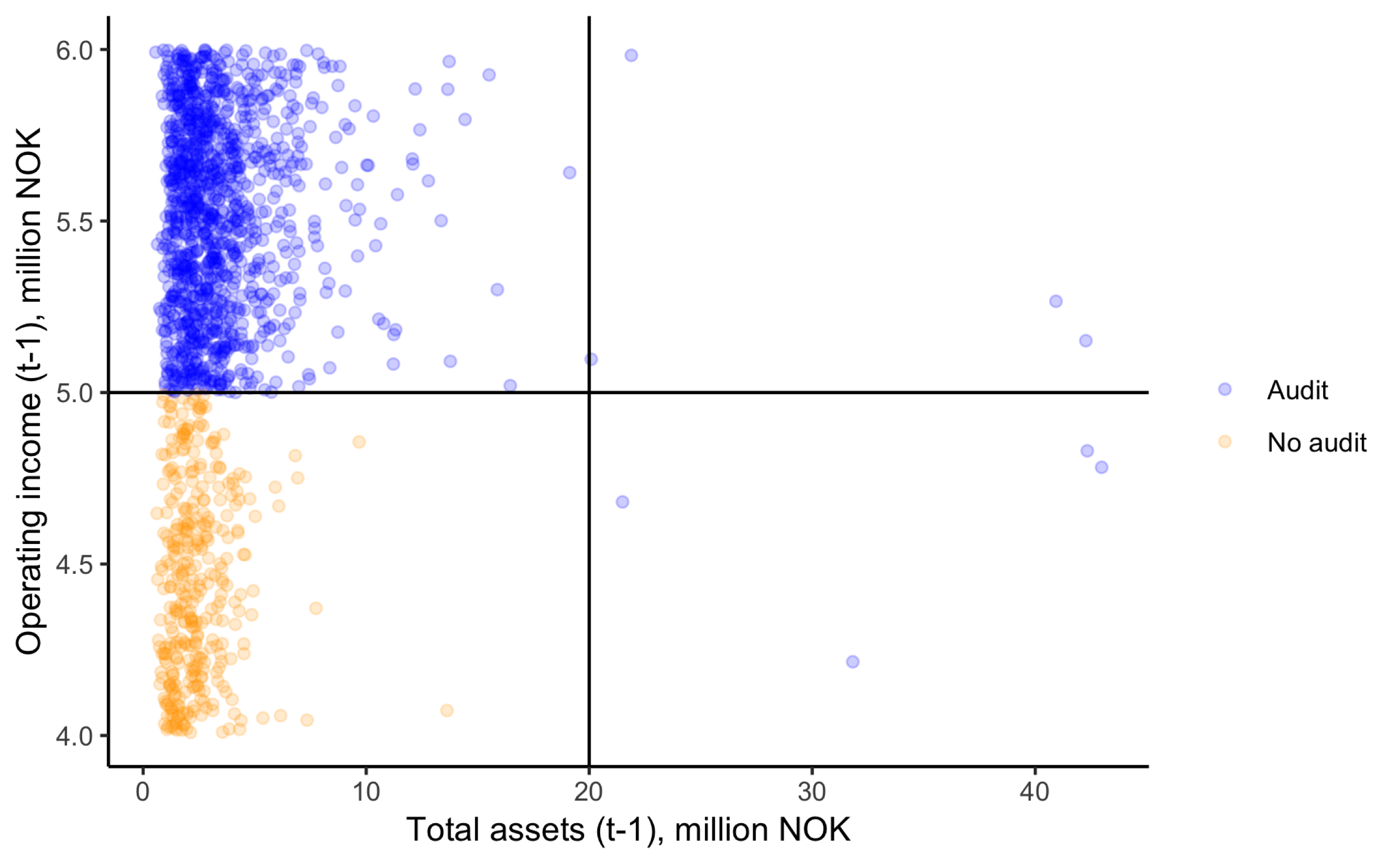}
 \end{minipage}
\bigskip

\end{figure}

In the fifth column, results are shown where the sample is limited to firms with operating revenue of between NOK 4 and 6 million, where the threshold for eligibility was 5 million. Figure \ref{fig:narrow_plot} shows a scatter plot of the full sample in the left panel and the narrow subset of data in the right panel, with operating revenue limited to be within plus or minus NOK 1 million of the threshold. By limiting the sample to firms closer to the threshold we are attempting to strengthen the ignorability assumption--that is, the assumption that assignment into either the control or treatment group is essentially random due to the arbitrariness of the threshold values. While the sample is reduced to a $\frac{1}{10}$th of the original, the estimated treatment effect remains approximately the same and is significant at the 1\% level.

\begin{table}[!htbp] \centering
  \caption{\label{tbl:discon} Results from the regression discontinuity design. The first column shows results with only an intercept, year fixed effects and the treatment indicator. The second column shows results that include covariates for firm size, risk, leverage and historical measures of cash flow. The coefficient on the \emph{noAudit} indicator is now positive and significant. From the third column, results are shown where firms who were eligible to forgo an auditor, but still chose to get audited were excluded. The fourth column shows results where 94 sector fixed effects are added (coefficients not reported). In the fifth column, only firms with operating revenue of between 4 and 6 million are included, where 5 million is the threshold for the exemption policy.}
\begin{tabular}{lccccc}
    &     I     &     II    &    III    &     IV    &     V      \\
\midrule
 & \multicolumn{5}{c}{\textit{Dependent variable:}} \\
\cline{2-6}
\\[-1.8ex] & \multicolumn{5}{c}{Dividend} \\
\hline \\[-1.8ex]
Intercept & $-$0.020 & $-$0.046$^{***}$ & $-$0.135$^{***}$ & $-$0.358$^{***}$ & $-$0.471 \\
 & (0.019) & (0.016) & (0.024) & (0.108) & (0.324) \\

noAudit & $-$0.275$^{***}$ & 0.044$^{*}$ & 0.152$^{***}$ & 0.140$^{***}$ & 0.161$^{***}$ \\
 & (0.025) & (0.023) & (0.030) & (0.030) & (0.047) \\

operating revenue &  & 0.048$^{***}$ & 0.094$^{***}$ & 0.101$^{***}$ & 0.295$^{***}$ \\
 &  & (0.009) & (0.013) & (0.013) & (0.110) \\

operating income (t-1) &  & 0.180$^{***}$ & 0.191$^{***}$ & 0.183$^{***}$ & 0.081$^{***}$ \\
 &  & (0.011) & (0.012) & (0.013) & (0.026) \\

employees &  & 0.014$^{*}$ & 0.019$^{**}$ & 0.005 & $-$0.046$^{*}$ \\
 &  & (0.008) & (0.009) & (0.010) & (0.027) \\

total assets &  & $-$0.021$^{**}$ & $-$0.038$^{***}$ & $-$0.043$^{***}$ & 0.078$^{***}$ \\
 &  & (0.009) & (0.011) & (0.011) & (0.020) \\

risk (roa sd) &  & $-$0.043$^{***}$ & $-$0.075$^{***}$ & $-$0.071$^{***}$ & $-$0.00003 \\
 &  & (0.009) & (0.013) & (0.013) & (0.024) \\

leverage &  & 0.107$^{***}$ & 0.147$^{***}$ & 0.150$^{***}$ & 0.035$^{***}$ \\
 &  & (0.008) & (0.010) & (0.011) & (0.012) \\

cash flow (mean) &  & 0.395$^{***}$ & 0.405$^{***}$ & 0.415$^{***}$ & 0.353$^{***}$ \\
 &  & (0.011) & (0.012) & (0.013) & (0.030) \\

cash (sd) &  & 0.054$^{***}$ & 0.065$^{***}$ & 0.052$^{***}$ & $-$0.018 \\
 &  & (0.009) & (0.011) & (0.011) & (0.021) \\

\midrule
Year FE         & YES & YES & YES & YES & YES \\
Sector FE 			& NO & NO & NO & YES & YES \\
R$^{2}$ &  & 0.325 & 0.341 & 0.351 & 0.218 \\
Adjusted R$^{2}$ &  & 0.325 & 0.340 & 0.347 & 0.192 \\
Residual Std. Error &  & 0.822 & 0.882  & 0.877  & 0.641 \\
N                 & 14330     & 14330     & 11110     & 11110     & 1837      \\
\bottomrule
\hline \\[-1.8ex]
 &  \multicolumn{5}{r}{\textit{Standard errors are adjusted for clustering}} \\
  & \multicolumn{5}{r}{\textit{$^{*}$p$<$0.1; $^{**}$p$<$0.05; $^{***}$p$<$0.01}} \\
\end{tabular}
\end{table}

\begin{figure}
\caption{The figure shows an illustration of the regression results, transformed back into the scale of NOK. The results are from a regression that only includes firms with operating revenue between 4 and 6 million kroner, where 5 million NOK is the threshold for forgoing an auditor. Firms that chose to be audited despite being eligible for the exemption are also discarded. The black lines represent the maximum likelihood estimate of the relationship between operating revenue and dividend. The estimated treatment effect is the gap between the black lines at the threshold. The grey lines represent uncertainty of the intercept term, the coefficient on operating revenue and the treatment effect. The uncertainty is in the form of draws from normal distributions with means and standard errors corresponding to the maximum likelihood estimations of the model. All other covariates are held fixed at their mean value.}
\centering
\includegraphics[width=.8\textwidth]{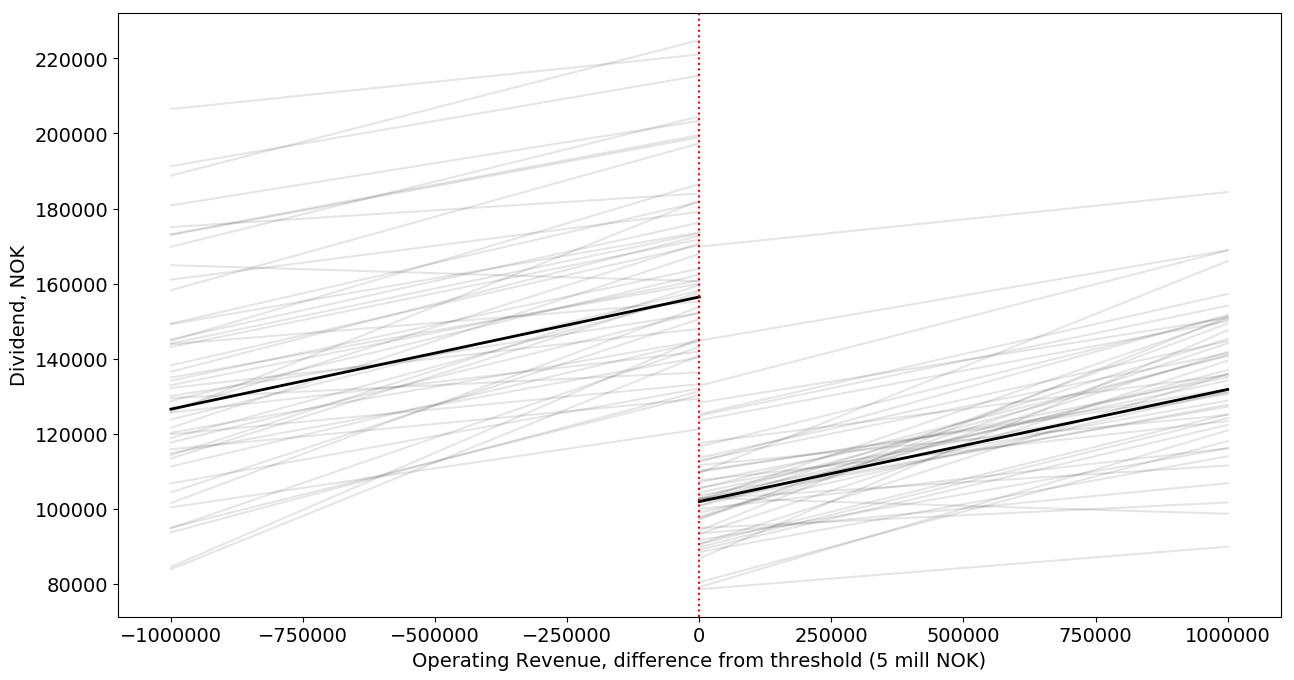}
\label{fig:reg_disc_figure}
\end{figure}

A visualization of the main results from the specification from column five in table 4 can be seen in figure \ref{fig:reg_disc_figure}. The black lines represent the maximum likelihood estimate of the relationship between operating revenue and dividends, where the discontinuity at the 5 million NOK mark (red dotted line) represents the estimated effect of the auditing exemption policy. The grey lines represent uncertainty, in the form of sampling variance, of the intercept term, the coefficient on operating revenue, and the treatment effect. The uncertainty is in the form of draws from normal distributions with means and standard errors corresponding to the maximum likelihood estimations of the model. All other covariates are held fixed at their mean value.

A simple robustness check for the regression discontinuity design is to run regressions with thresholds that are immaterial. We run two such checks. First, we change the threshold from the actual 5 million NOK threshold for operating revenue to 7 million. In another check, we identify the firms that were just under the threshold and did not receive an audit but use data from 2010, before the implementation of the policy. In both cases we limited firms to those that were within 1 million NOK of the operating revenue threshold. In both cases, the threshold indicator is estimated to be close to zero and not significant. The results are shown in appendix A.

We hypothesize that auditing reduces dividends through conservative accounting practices constraining current earnings. We provide direct evidence of this mechanism by running a regression where earnings are the dependent variable. Table \ref{tbl:earningsResults} shows the results. The first column shows results with both year and sector fixed effects. As in the results from columns three through five in table \ref{tbl:discon}, firms who were eligible to forgo an auditor, but still chose to get audited are excluded. The coefficient on the \emph{noAudit} indicator is positive and both statistically and economically significant .24 standard deviations. We can interpret this as firms that were able to forgo an audit increasing their reported earnings by approximately 160,000 NOK (€16,000).

\begin{table}[!htbp] \centering
  \caption{Earnings as the dependent variable. Column one shows results with both year fixed effects and sector fixed effects. Firms who were eligible to forgo an auditor, but still chose to get audited are excluded. In the second column, only firms with operating revenue of between 4 and 6 million are included. Firms who were eligible to forgo an auditor, but still chose to get audited are excluded.}
  \label{tbl:earningsResults}
\begin{tabular}{@{\extracolsep{5pt}}lcc}
\\[-1.8ex]\hline
\hline \\[-1.8ex]
 & \multicolumn{2}{c}{\textit{Dependent variable:}} \\
\cline{2-3}
\\[-1.8ex] & \multicolumn{2}{c}{Earnings} \\
\\[-1.8ex] & (1) & (2)\\
\hline \\[-1.8ex]
 Intercept & 0.348$^{***}$ & 0.348$^{***}$ \\
  & (0.022) & (0.022) \\
  & & \\
 noAudit & 0.240$^{***}$ & 0.240$^{***}$ \\
  & (0.010) & (0.010) \\
  & & \\
 operating revenue & 0.204$^{***}$ & 0.204$^{***}$ \\
  & (0.009) & (0.009) \\
  & & \\
 operating income (t-1) & $-$0.018$^{**}$ & $-$0.018$^{**}$ \\
  & (0.007) & (0.007) \\
  & & \\
 employees & 0.087$^{***}$ & 0.087$^{***}$ \\
  & (0.008) & (0.008) \\
  & & \\
 total assets & $-$0.034$^{***}$ & $-$0.034$^{***}$ \\
  & (0.010) & (0.010) \\
  & & \\
 risk (roa sd) & $-$0.021$^{***}$ & $-$0.021$^{***}$ \\
  & (0.008) & (0.008) \\
  & & \\
 leverage & 0.531$^{***}$ & 0.531$^{***}$ \\
  & (0.009) & (0.009) \\
  & & \\
 cash flow (mean) & 0.008 & 0.008 \\
  & (0.008) & (0.008) \\
  & & \\
 cash (sd) & $-$0.039 & $-$0.039 \\
  & (0.079) & (0.079) \\
  & & \\
\hline \\[-1.8ex]
Observations & 11,110 & 11,110 \\
R$^{2}$ & 0.644 & 0.644 \\
Adjusted R$^{2}$ & 0.642 & 0.642 \\
Residual Std. Error (df = 11040) & 0.642 & 0.642 \\
\hline
\hline \\[-1.8ex]
\textit{Note:}  & \multicolumn{2}{r}{$^{*}$p$<$0.1; $^{**}$p$<$0.05; $^{***}$p$<$0.01} \\
\end{tabular}
\end{table}

\begin{figure}
\centering
\caption{Estimated coefficients on the audit exemption indicator for each year 2011 through 2015. The effect appears roughly constant throughout the period, with no indication of a reduction over time.}
\label{fig:time_effect_plot}
\includegraphics[width=.6\textwidth]{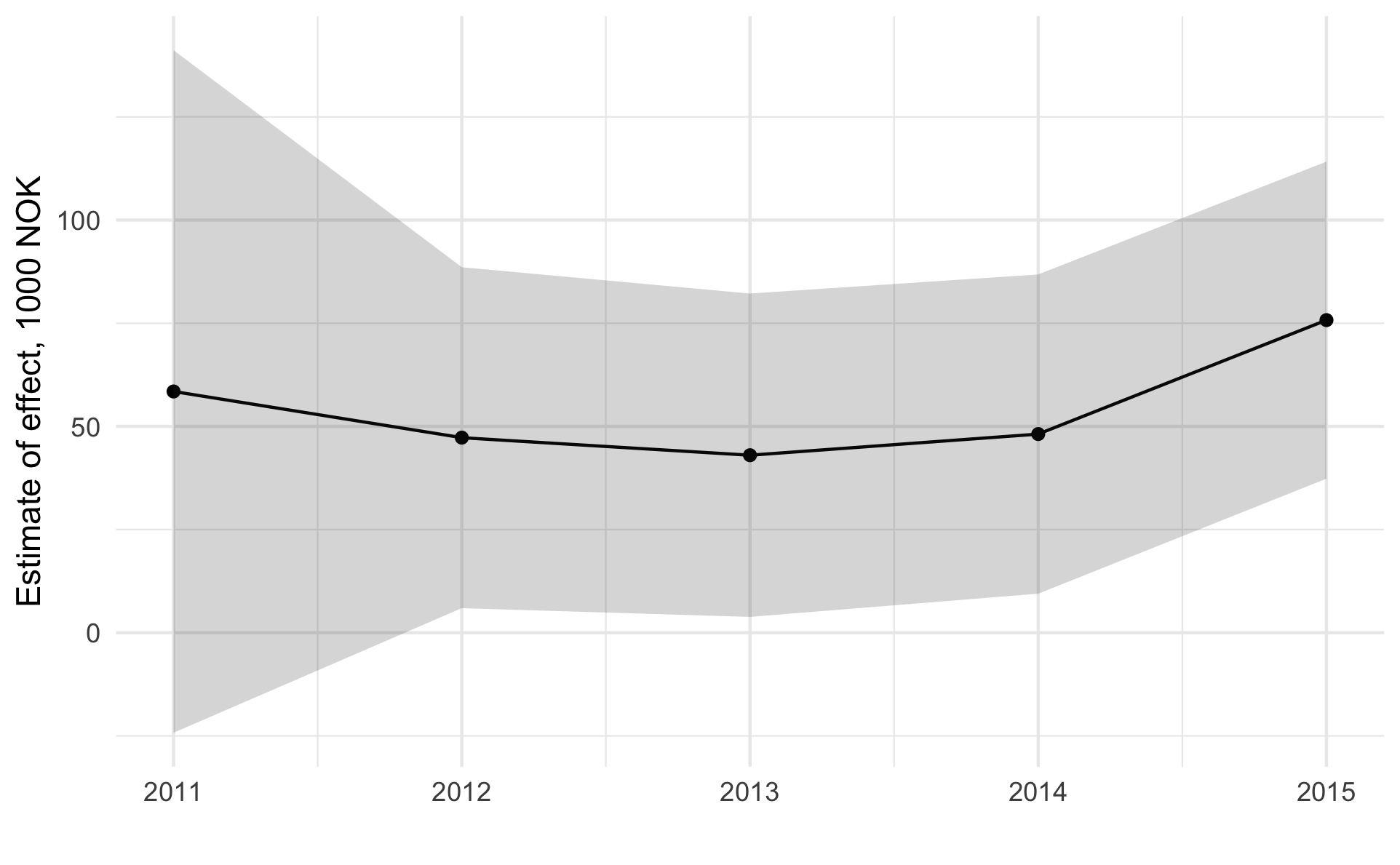}

\end{figure}

Plausibly, the effect of an audit could be short term. As firms adjust to the audit exemption, dividends may return to a normal level. We only have data through 2015--four years after the initial roll-out of the policy. But we can attempt to address this question by estimating separate effects for the policy for each of the five years 2011-2015. Results for these regressions are shown in table \ref{tbl:time_effects}. The three columns show results from regressions where firms who were eligible to forgo an audit, but still chose to get audited were excluded. The second column shows results from a specification with sector fixed effects. The third column only includes firms with operating revenue between 4 and 6 million. The main results are presented in figure \ref{fig:time_effect_plot}. The figure represents the point estimates and confidence intervals for coefficients on the audit exemption variable (\emph{noAudit}) for each year 2011-2015, transformed back into the original units of 1000 NOK. The estimated effect remains roughly constant throughout the period.

\begin{table}[!htbp] \centering
  \caption{Results from the regression discontinuity analysis where the effect of forgoing an audit is estimated each year from 2011 through 2015. Firms who were eligible to forgo an auditor, but still chose to get audited are excluded. The second column shows results where 94 sector fixed effects are added (coefficients not reported). In the third column, only firms with operating revenue of between 4 and 6 million are included, where 5 million is the threshold for participation in the reform. The effect of forgoing an auditor appears relatively constant through the years 2011 to 2015, with no indication of drop-off after the initial roll-out of the policy.}
  \label{tbl:time_effects}
\begin{tabular}{@{\extracolsep{5pt}}lccc}
\\[-1.8ex]\hline
\hline \\[-1.8ex]
 & \multicolumn{3}{c}{\textit{Dependent variable:}} \\
\cline{2-4}
\\[-1.8ex] & \multicolumn{3}{c}{Dividend} \\
\\[-1.8ex] & I & II & III\\
\hline \\[-1.8ex]
intercept & $-$0.147$^{***}$ & $-$0.339$^{***}$ & $-$0.436 \\
  & (0.024) & (0.109) & (0.325) \\
noAudit:2011 & 0.167 & 0.150 & 0.015 \\
  & (0.106) & (0.106) & (0.180) \\
noAudit:2012 & 0.141$^{***}$ & 0.121$^{**}$ & 0.137 \\
  & (0.053) & (0.053) & (0.086) \\
noAudit:2013 & 0.127$^{**}$ & 0.110$^{**}$ & 0.151$^{*}$ \\
  & (0.050) & (0.050) & (0.082) \\
noAudit:2014 & 0.135$^{***}$ & 0.123$^{**}$ & 0.159$^{**}$ \\
  & (0.050) & (0.050) & (0.080) \\
noAudit:2015 & 0.208$^{***}$ & 0.194$^{***}$ & 0.210$^{**}$ \\
  & (0.049) & (0.049) & (0.083) \\
operating revenue & 0.099$^{***}$ & 0.106$^{***}$ & 0.284$^{**}$ \\
  & (0.013) & (0.014) & (0.110) \\
employees & 0.022$^{**}$ & 0.007 & $-$0.045$^{*}$ \\
  & (0.009) & (0.010) & (0.027) \\
total assets & $-$0.014 & $-$0.021$^{*}$ & 0.084$^{***}$ \\
  & (0.011) & (0.011) & (0.020) \\
risk (sd roa) & $-$0.083$^{***}$ & $-$0.078$^{***}$ & $-$0.010 \\
  & (0.013) & (0.014) & (0.024) \\
leverage & 0.140$^{***}$ & 0.146$^{***}$ & 0.034$^{***}$ \\
  & (0.011) & (0.011) & (0.012) \\
cash flow (mean) & 0.537$^{***}$ & 0.544$^{***}$ & 0.405$^{***}$ \\
  & (0.009) & (0.009) & (0.024) \\
cash flow (sd) & 0.079$^{***}$ & 0.064$^{***}$ & $-$0.008 \\
  & (0.011) & (0.011) & (0.021) \\
\hline \\[-1.8ex]
Sector FE & NO & YES & YES \\
N & 11,110 & 11,110 & 1,839 \\
R2 & 0.327 & 0.339 & 0.214 \\
Adjusted R2 & 0.326 & 0.334 & 0.186 \\
\hline
\hline \\[-1.8ex]
\textit{Note:}  &  \multicolumn{3}{r}{\textit{Standard errors are adjusted for clustering}} \\
& \multicolumn{3}{r}{\textit{$^{*}$p$<$0.1; $^{**}$p$<$0.05; $^{***}$p$<$0.01}} \\
\end{tabular}
\end{table}

A potential weakness of the above analysis is that the relationship between dividend payouts and firm size--as represented by operating revenue--is assumed to be linear. Plausibly, a non-linear relationship between dividend payout and firm size could lead to a biased estimate of the discontinuity term. A common solution is to model the continuous forcing variable, in our case operating revenue, as a polynomial. However, \citet{gelman_why_2017} warn that use of polynomials in the forcing variable can lead to noisy estimates and results that are sensitive to the order of the polynomial. They instead recommend the use of smoothed functions.

\begin{figure}
\centering
\caption{The figure shows predictions from the first stage regression, where separate smoothed functions of operating revenue are fit to the data below and above the 5 million NOK threshold.}
\includegraphics[width=.6\textwidth]{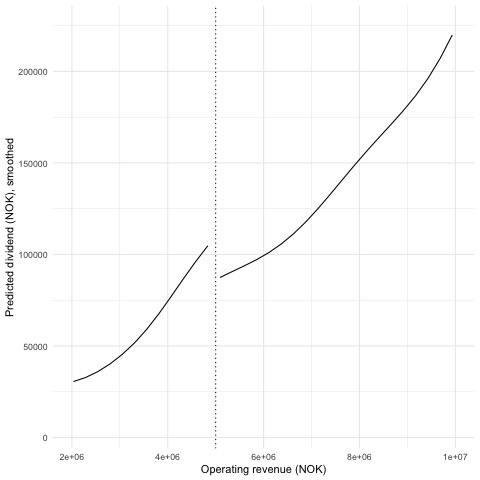}

\label{fig:smoothed_curves}
\end{figure}

We therefor use a two-step regression strategy that first estimates separate smooth functions of operating revenue on dividends, over and under the policy cut-off. The smooth functions are estimated as penalized cubic regression splines where the smoothing parameter is chosen optimally through cross validation. For technical details see \citet{hastie_generalized_1990} and  \citet{wood_generalized_2006, wood_mgcv:_2001}. Figure \ref{fig:smoothed_curves} shows predictions from the first stage regression. The overall relationship between dividends and operating revenue is positive, as would be expected. The discontinuity at the policy cut-off point is also clear.

In the second stage, the residuals from the first stage regression are used as the dependent variable in a linear regression of the \emph{noAudit} indicator and control variables.

Starting from equation \ref{eqn:reg_disc}, we can re-write the regression discontinuity as in equation \ref{eqn:reg_disc_smoothed}, where $f(\mathbf{SIZE})_{s=audit}$ and $f(\mathbf{SIZE})_{s=exempt}$ represent the separate smoothed functions of the size indicator, operating revenue, and $noAudit_{it}$ represents the discontinuity estimate with the use of a smoothed function of the forcing variable. The vector of controlling covariates is again represented by $\mathbf{X\beta}$.

\begin{equation}
DIV_{it}=\tau noAudit_{it}^{smooth}  + f(\mathbf{SIZE})_{s=audit} + f(\mathbf{SIZE})_{s=exempt} +  \mathbf{X \beta} + \epsilon_{it}
\label{eqn:reg_disc_smoothed}
\end{equation}

Equation \ref{eqn:two-stage} shows the component equations of the two-stage regression.

\begin{align}
\begin{split}
    DIV_{it} &= f(\mathbf{SIZE})_{s=audit} + f(\mathbf{SIZE})_{s=exempt} + \epsilon_{it}^{stage=1} \\
    \hat{\epsilon^{stage=1}} &= NoAudit_{it}^{smooth}  + \mathbf{X \beta} + \epsilon_{it}^{stage=2}
    \label{eqn:two-stage}
\end{split}
\end{align}

Results from the two-stage regression are presented in table \ref{tbl:two_stage}. Standard errors and p-value indicators are adjusted to account for the full uncertainty in the two-stage regression. In the first column, the entire sample is included in the regression, including firms that voluntarily obtained an audit. The audit exemption variable becomes positive and statistically significant. In columns 2-4, firms that chose to receive an audit despite not being required to are excluded. The coefficient on the audit-exemption variable is estimated at a magnitude that is larger than in the model with a linear representation of operating revenue. When sector fixed effects are added (fourth column), the results are not materially changed. When the analysis is limited to firms with operating revenue between 4 and 6 million NOK (column 5) the estimated coefficient is reduced in magnitude but still significant and positive.

The point estimate of approximately 0.23 on the \emph{noAudit} indicator in the 3rd and 4th columns can be interpreted as an audit exemption leading to a .23 standard deviation increase in a firm's dividend. This corresponds to approximately 85,000 NOK (approximately €9,000), an economically significant amount for a small firm. Even the smaller estimate from the narrow estimation in column 5, with a coefficient of .08 corresponds to approximately 35,000 NOK (approximately €4,000). This range of estimates is roughly in line with the linear models presented earlier.

\begin{table}[!htbp] \centering
  \caption{Results from a two-stage estimation with a smoothed function of firm operating revenue. The first column includes the entire sample, including firms who voluntarily obtained an audit. In the second column, firms who voluntarily chose to obtain an audit are dropped. In the third column, sector fixed effects are included. In the fourth column, only firms with operating revenue between 4 and 6 million NOK are included in the analysis.}
  \label{tbl:two_stage}
\begin{tabular}{lccccc}
\\[-1.8ex]\hline
\hline \\[-1.8ex]
 & \multicolumn{4}{c}{\textit{Dependent variable}} \\
\cline{2-5}
\\[-1.8ex]  & \multicolumn{4}{c}{Dividend} \\
\\[-1.8ex]  & (1) & (2) & (3) & (4)\\
\hline \\[-1.8ex]
intercept & $-$0.019 & $-$0.120$^{***}$ & $-$0.285$^{***}$ & $-$0.252 \\
& (0.016) & (0.021) & (0.109) & (0.324) \\
 noAudit  & 0.151$^{***}$ & 0.257$^{***}$ & 0.225$^{***}$ & 0.083$^{**}$ \\
  & (0.022) & (0.025) & (0.026) & (0.040) \\

 employees   & $-$0.036$^{***}$ & $-$0.011 & $-$0.031$^{***}$ & $-$0.041 \\
   & (0.007) & (0.008) & (0.010) & (0.027) \\

 risk (sd roa)   & $-$0.021$^{**}$ & $-$0.062$^{***}$ & $-$0.060$^{***}$ & $-$0.011 \\
    & (0.009) & (0.013) & (0.014) & (0.024) \\

 leverage   & 0.070$^{***}$ & 0.115$^{***}$ & 0.123$^{***}$ & 0.034$^{***}$ \\
    & (0.008) & (0.011) & (0.011) & (0.012) \\

 cash flow (mean)   & 0.484$^{***}$ & 0.516$^{***}$ & 0.524$^{***}$ & 0.407$^{***}$ \\
    & (0.008) & (0.009) & (0.009) & (0.024) \\

cash flow (sd)   & 0.032$^{***}$ & 0.061$^{***}$ & 0.046$^{***}$ & $-$0.009 \\
    & (0.009) & (0.011) & (0.011) & (0.021) \\

\hline \\[-1.8ex]
Year FE  & YES & YES & YES & YES \\
Sector FE 		& NO & NO &  YES & YES \\
N  & 14,330 & 11,110 & 11,110 & 1,837 \\
R$^{2}$ & 0.247 & 0.275 & 0.289 & 0.210 \\
Adjusted R$^{2}$ & 0.246 & 0.275 & 0.284 & 0.185 \\
\hline
\hline \\[-1.8ex]
\textit{Note: }  & \multicolumn{4}{r}{$^{*}$p$<$0.1; $^{**}$p$<$0.05; $^{***}$p$<$0.01} \\
\end{tabular}
\end{table}

\FloatBarrier

%\subsection{Directly modeling heterogeneity in the regression discontinuity: mixed effects models}

\section{Regression discontinuity with matched control \label{matched}}

The main empirical challenge in this article is that the correct control group--firms that were not eligible to forgo an auditor but would have forgone an auditor if they had the opportunity--is not observed. In the previous section, we have attempted to correct for this potential source of bias by including linear control variables in our estimation. We have also introduced a non-parametric representation of the forcing variable--operating revenue--to take account of the potential for a non-linear relationship with dividend policy. Still, such controls only partially correct for imbalance between treatment and control groups \citep{imbens_causal_2015}.

In this section we attempt to directly improve the balance between the treatment and control groups using propensity score matching \citep{rubin_using_1979, imbens_causal_2015}. Intuitively, we are seeking to create an artificial control group that most resembles the treatment group.

We start by creating a logit regression model with all firms that were eligible to forgo an audit, where the dependent variable is whether they chose to forgo auditing (1) or not (0). This can be represented simply as in equation \ref{eqn:noaudit}, where $\mathbf{X}$ represents the matrix of predictor terms. The predicted log-odds from this model, $logit(\hat{Y_{noaudit}})$, which we will refer to as $\hat{l}$, are the propensity scores.

\begin{equation}
logit(Y_{noaudit}) = \mathbf{X} \beta
\label{eqn:noaudit}
\end{equation}

In the matrix of predictors for the propensity score model, we include the continuous variables operating revenue, employees, total assets, leverage, return on assets and the mean and standard deviation of cash flow. we also include squared terms for all these variables as well as interaction effects between employees and operating revenue. We include fixed effects for years and industry sector. The model is complex, but the main purpose is not descriptive or interpretive, but rather to create the best possible balance between treatment and control.

\begin{figure}[ht]
\caption{Balanced control. The figure illustrates the balance of the sample between the treatment and control groups before and after matching. The top panel shows a histogram of the propensity scores of the treatment group: The firms that were eligible to forgo an auditor and chose to forgo an auditor. The middle column shows the histogram of the propensity scores of the control group: the firms in our sample that were not eligible to forgo auditing. The balance of the sample is poor. In the third column, the histogram of the control group after matching is shown.}
\centering
\includegraphics[width=.8\textwidth]{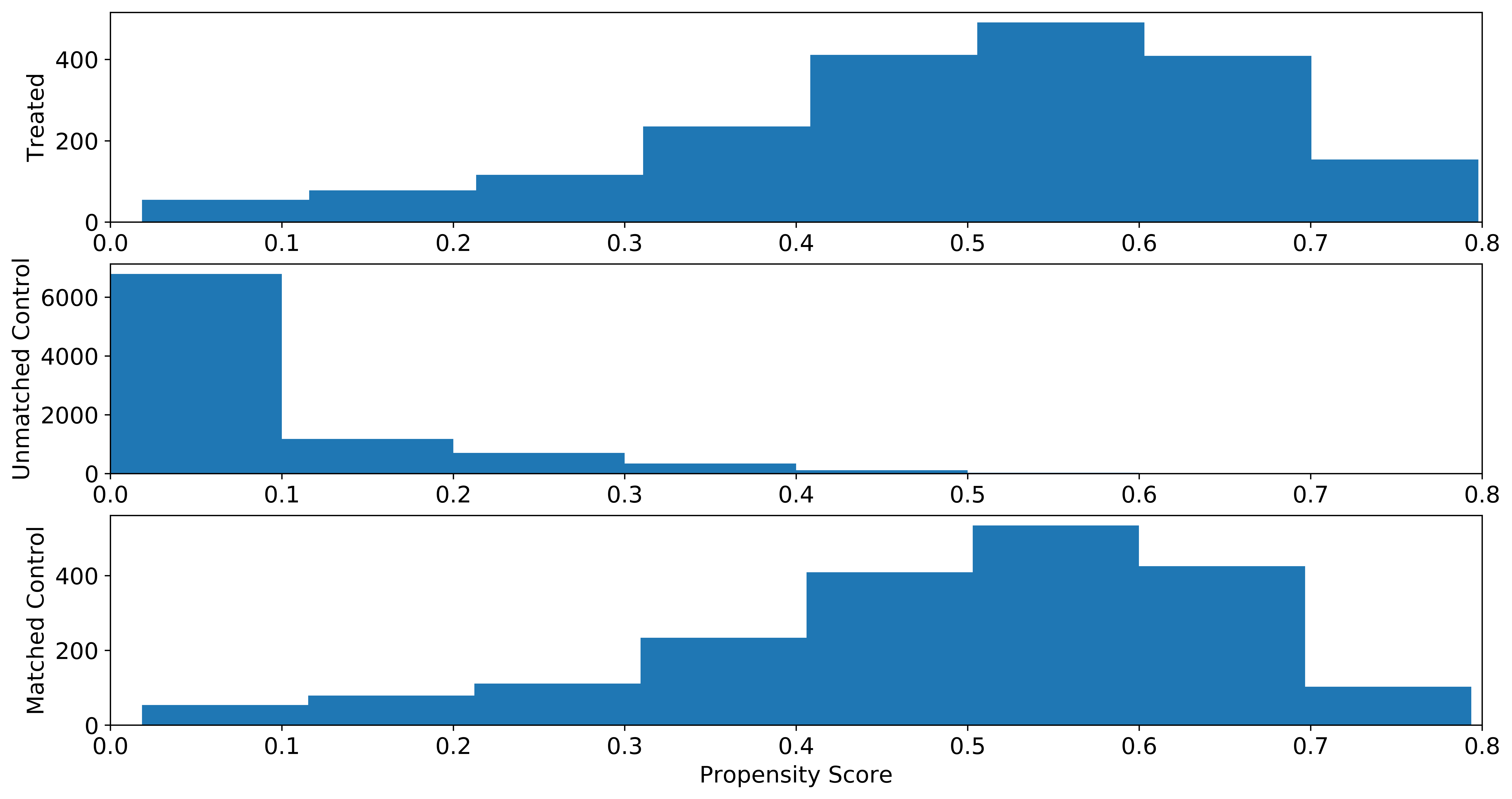}
\label{fig:matched_balance}
\end{figure}

From the logit model, we create the propensity scores for the treatment group, $\hat{l}_{treat}$, consisting of the observations with firms under the threshold who chose to forgo auditing. A histogram of the treatment propensity scores is shown in the top panel of figure \ref{fig:matched_balance}.  We also create propensity scores for the sample of potential controls--firms not eligible to forgo auditing--that we can designate $\hat{l}_{unmatched}$. A histogram of these propensity scores is shown in the middle panel of figure \ref{fig:matched_balance}. The figure shows that the treatment and unmatched control groups are poorly balanced. We then use a nearest-neighbor matching algorithm so that for each treatment observation, a control observation with propensity score, $\hat{l}_{con}$ is chosen such that the squared difference, $(\hat{l}_{treat}-\hat{l}_{con})^2$, is minimized. A histogram of the matched control sample is shown in the bottom panel of figure \ref{fig:matched_balance}. The balance between the treatment and control is now improved.

With a balanced control, we proceed with the regression analysis. The results are presented in table \ref{tbl:matching}. In the first column, only the \emph{noAudit} indicator and year fixed effects are included in the regression. This regression can be seen as a simple comparison of means in the treatment and control groups. The coefficient on the  \emph{noAudit} indicator is estimated to be positive and significant. In the second column, we add operating revenue as a control variable. We can now interpret the treatment indicator at the policy threshold for operating revenue, rather than just comparing means between groups. The estimated coefficient on the treatment indicator is now nearly doubled to 0.109 standard deviations. In the third column we also add the other continuous variables as controls and in the fourth column we add sector fixed effects. In the fifth column we limit the sample to firms with operating revenue within one million NOK of the threshold. The estimated coefficient on the treatment indicator, \emph{noAudit}, remains in the range of 0.10 to 0.16 standard deviations. Converted back to the original units, this corresponds to between approximately 40,000 and 60,000 NOK (€4,000-6000).

\begin{table}[ht]
\caption{Results from the regression discontinuity design with matched control group. The first column shows results with only the treatment indicator (\emph{noAudit}) and year fixed effects. The coefficient on the treatment indicator is positive and significant. In the second column we add operating revenue to the regression so that the coefficient on the \emph{noAudit} indicator is interpreted at the threshold. In the third column we add other control variables. In the fourth column we add sector fixed effects and in the fifth column we limit the sample to firms with operating revenue within one million NOK of the voluntary auditing threshold of 5 million NOK. The coefficient on the \emph{noAudit} indicator remains significantly positive in the range of 0.10 to 0.16 standard deviations. \label{tbl:matching}}
\begin{center}
\begin{tabular}{lccccc}
\\[-1.8ex]\hline
\hline \\[-1.8ex]
 & \multicolumn{5}{c}{\textit{Dependent variable:}} \\
\cline{2-6}
 & \multicolumn{5}{c}{\textit{Dividend}} \\
                    &     I     &     II    &    III    &     IV & V   \\
\midrule
Intercept          &  -0.300*** &  -0.302*** &  -0.190*** &  -0.152*** &    -0.123 \\
                   &    (0.030) &    (0.029) &    (0.030) &    (0.040) &   (0.627) \\
noAudit            &   0.056*** &   0.104*** &   0.130*** &   0.107*** &  0.160*** \\
                   &    (0.015) &    (0.019) &    (0.018) &    (0.020) &   (0.043) \\
operating revenue    &            &   0.074*** &   0.132*** &   0.106*** &    0.145* \\
                   &            &    (0.013) &    (0.019) &    (0.023) &   (0.087) \\
employees     &            &            &  -0.014*** &     -0.012 &    -0.000 \\
                   &            &            &    (0.003) &    (0.008) &   (0.016) \\
total assets          &            &            &   0.095*** &     0.080* &     0.182 \\
                   &            &            &    (0.029) &    (0.046) &   (0.235) \\
risk (roa sd)           &            &            &      0.013 &      0.006 &     0.080 \\
                   &            &            &    (0.019) &    (0.021) &   (0.110) \\
leverage           &            &            &    0.029** &    0.031** &    0.026* \\
                   &            &            &    (0.013) &    (0.014) &   (0.015) \\
cash flow (mean)       &            &            &   0.350*** &   0.405*** &  0.682*** \\
                   &            &            &    (0.041) &    (0.047) &   (0.112) \\
cash flow (sd)       &            &            &  -0.100*** &  -0.095*** &    -0.198 \\
                   &            &            &    (0.022) &    (0.024) &   (0.134) \\
\midrule
Year FE            &        YES & YES & YES & YES & YES\\
Sector FE          &        NO  &         NO &       YES    &     YES   &   YES \\
N                  &       3902 &       3902 &       3902 &       3902 &      1341 \\
R2                 &       0.01 &       0.02 &       0.14 &       0.18 &      0.26 \\
Adjusted R2        &       0.01 &       0.02 &       0.14 &       0.16 &      0.23 \\
\bottomrule
\textit{Note: }  &  \multicolumn{5}{r}{Standard errors are adjusted for clustering} \\
\textit{}  & \multicolumn{5}{r}{$^{*}$p$<$0.1; $^{**}$p$<$0.05; $^{***}$p$<$0.01} \\
\end{tabular}
\end{center}

\end{table}

\FloatBarrier

\section{Difference-in-difference estimators}

The regression discontinuity design takes advantage of the arbitrariness of the policy thresholds on firm size. However, the design does not make full use of the panel of data to inform the estimates, as only data after the introduction of the policy is used.

As a robustness check, we make use of data prior to the introduction of the policy to do a difference-in-difference estimation. The basic intuition is that by comparing not just between treatment and control groups but also before and after the introduction of the policy, the effect of unobserved time-invariant covariates that affects selection into the treatment group can be controlled for.

The results from the difference-in-difference estimation are in line with the results from the regression discontinuity design. Coefficients on the \emph{noAudit} indicator are estimated to be between .12 and .20, though these results are estimated with higher standard errors due to the inefficiency introduced due to differencing. Details of the difference-in-difference estimation and a full presentation of results can be found in Appendix C.

\FloatBarrier
\section{Conclusion}

%The auditor has a vital role in modern economies by serving as an informational intermediary between firms and outside investors, regulators, and other stakeholders. In addition, auditors can promote financial conservatism and appropriate accounting and control practices.

%Identifying the effect of auditing on firm behavior is challenging and is an active area of research. Research on non-listed firms is limited due to scarce data. A challenge when working with data of publicly traded firms is that auditing is nearly universally required. Researchers must then attempt to estimate differences between grades of audit quality. This is only indirectly measured through proxy variables which may both introduce measurement error into estimation as well as issues of endogeneity.

The level of dividends is a key decision that managers of small private firms make. Dividends reward owners for the capital they have invested. In the case of small firms where owners often serve as managers, dividends also reward the large expenditures of time and effort that owners often put into their firms. Yet managers and owners of small firms may lack financial and accounting sophistication and may not be fully aware of conservative accounting principles.

Auditing can have the effect of promoting the consistent application of conservative accounting principles and imposing a degree of financial discipline. All else equals, this will have the effect of reducing current earnings and in turn constraining dividends.

% One of the most important decisions the manager of a small firms makes is how much cash to extract from the firm, that is the dividend.  We contribute to this important strand of literature by making use of rich data on small private firms in Norway and focusing on the recent introduction of an audit exemption policy in 2011. To establish a mechanism for random treatment assignment, we use eligibility thresholds that are considered arbitrary within a narrow range of firm size.

Our hypothesis is that forgoing an audit following the implementation of an audit exemption policy for small firms in Norway could lead to an increase in dividends, all else equal. Using a regression discontinuity design with both linear and non-parametric representations of the forcing variable, propensity score matching as well as difference-in-difference estimation, we consistently find statistically and economically significant effects of forgoing an audit. A representative firm on the eligibility threshold that could, and did forgo an audit, increased their average dividend by between 35-85,000 NOK (€4,000 - €9,000) compared to similar firms just above the threshold. This amount is greater than the average cost savings of forgoing an audit.

%We cannot definitively identify a mechanism for the effects we estimate with our available data. However, the results are consistent with a mechanism that is especially relevant to small and medium-sized firms. Such firms typically have owners and managers that lack formal training in accounting and finance. Auditors can then play a role of promoting accounting conservatism through their duty to report issues with the firms going concern. Conversations with practitioners confirm that auditors to small- and medium-sized firms often play this role. %Through this \emph{adult in the room} role, auditors could influence owners and managers to favor retaining earnings over dividend payouts.

The results presented in this paper have limitations. For one, the regression discontinuity design requires that results be interpreted at the discontinuity threshold. Extrapolating the results both towards much smaller or much larger firms is not necessarily directly supported by the analysis. On the other hand, we have ex-ante no reason to believe that the results should hold only at the threshold. The theoretical underpinnings for our analysis are not contingent on a particular firm size. Neither is there anything in our empirical study that would indicate that the findings are limited to firms of a certain size.

The results should not be extrapolated to the extremes of either very small, owner-employee firms or large publicly listed firms. Yet even if the results pertain primarily to small and medium sized firms, the results are nonetheless important. Small and medium sized firms make up sizeable portions of employment and turnover in most developed economies yet receive relatively little attention by researchers compared to large publicly traded firms. The results also contribute to the understanding of auditor exemption policies. Our results indicate that the implementation of the exemption policy in Norway lead to a change in real economic behavior among small firms who did not obtain an audit. Our research does not support a normative interpretation of this change in behavior, however, and we consider a judgement of the optimality of the audit exemption policy or any level of dividends as outside the scope of this article. This article also provides evidence supporting the role of auditors in promoting accounting conservatism, and in turn how accounting conservatism can have real economic effects on the behavior of small firms.

\FloatBarrier
\section{Declaration of interest statement}

The authors have no conflicts of interest to disclose.
\FloatBarrier

\bibliographystyle{chicago}
\bibliography{accounting_quality}

\FloatBarrier
\newpage

\section*{Appendix A: Additional Descriptive Information}

\begin{figure}
\centering
\caption{One-year lagged operating revenue is shown on the vertical axis while number of employees are shown on the horizontal axis. The horizontal red line represents the threshold for operating revenue, while the vertical red line represents the threshold for number of full-time equivalent employees. The blue dots represent firms that were audited, while yellow dots represent firms that were not audited. The position of the dots is slightly jittered in order to aid readability. The figure shows that the 10-employee threshold does not correspond well to the number of employees in the data as the employee account is per head.}
\label{fig:income_employee_scat}
\includegraphics[width=.8\textwidth]{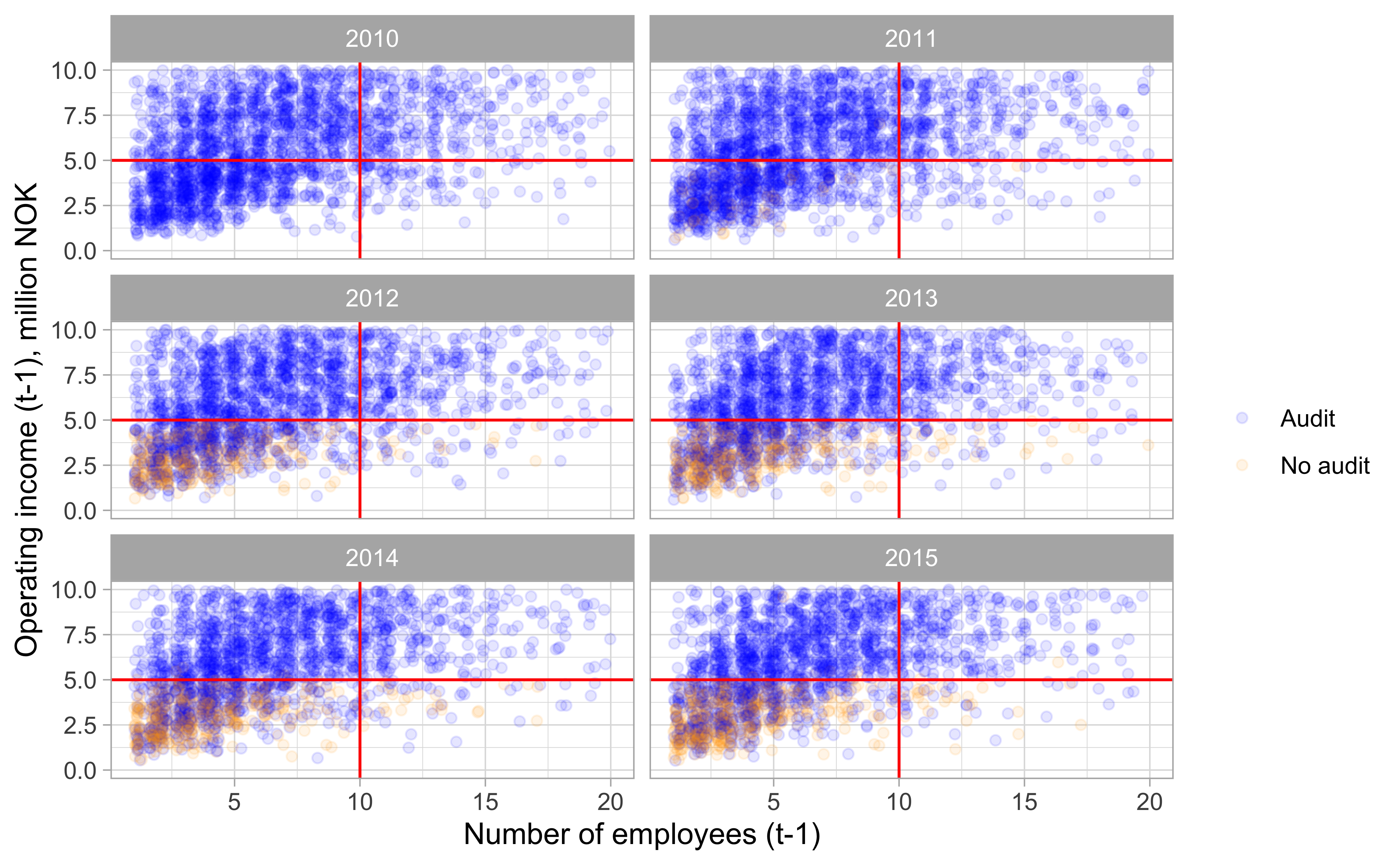}
\end{figure}

Figure \ref{fig:income_employee_scat} shows a scatter plot of 1-year lagged operating revenue and number of employees for firms in the sample. Here the red lines represent the thresholds for mandatory auditing. Similar to figure \ref{fig:scatter_size}, we see that operating revenue is the binding constraint for most firms. It is also apparent that the threshold for employees, which is established in terms of full-time equivalent employees does not match well with the employee data in the data set, which is per head. Thus two employees with 50\% positions would be counted as 2 employees in the data, but would count only as 1 employee relative to the threshold. The implications of this are that we can not reliably use the employee threshold as a discontinuity in our estimation. On the other hand, the threshold for employees is binding for even fewer firms than the figure initially indicates.

\FloatBarrier
\section*{Appendix B: Robustness checks, regression discontinuity}\label{Appendix A}

\begin{table}[!htbp] \centering
  \caption{Regression on firms who are eligible for audit exemption. The first column shows a regression with year fixed-effects but no other control variables. The second column shows a regression with control variables and sector fixed effects. The estimated coefficient on the noAudit indicator variables, which indicates whether a firm chose to forgo auditing is not estimated to be statistically significant}
  \label{tbl:underThreshold}
\begin{tabular}{@{\extracolsep{5pt}}lcc}
\\[-1.8ex]\hline
\hline \\[-1.8ex]
 & \multicolumn{2}{c}{\textit{Dependent variable:}} \\
\cline{2-3}
\\[-1.8ex] & \multicolumn{2}{c}{Dividends} \\
\\[-1.8ex] & \textit{normal} & \textit{OLS} \\
\\[-1.8ex] & (1) & (2)\\
\hline \\[-1.8ex]

Intercept & $-$0.203$^{***}$ & $-$0.100 \\
 & (0.017) & (0.106) \\

 noAudit & $-$0.023 & 0.023 \\
  & (0.017) & (0.016) \\

 operating revenue &  & 0.084$^{***}$ \\
  &  & (0.020) \\

employees &  & $-$0.026$^{*}$ \\
  &  & (0.015) \\

total assets&  & 0.104$^{***}$ \\
  &  & (0.014) \\

 risk (ROA) &  & $-$0.002 \\
  &  & (0.007) \\

 leverage &  & 0.030$^{***}$ \\
  &  & (0.006) \\

 cash flow (mean) &  & 0.334$^{***}$ \\
  &  & (0.015) \\

 cash flow (sd) &  & 0.009 \\
  &  & (0.014) \\
\hline \\[-1.8ex]
Year FE 			&      YES &        YES \\
Sector FE 			&      NO &        YES \\
R$^{2}$ &  & 0.164 \\
Adjusted R$^{2}$ &  & 0.153 \\
N & 5171 & 5171 \\
Residual Std. Error &  & 0.509\\
\hline
\hline \\[-1.8ex]
\textit{Note:}  & \multicolumn{2}{r}{$^{*}$p$<$0.1; $^{**}$p$<$0.05; $^{***}$p$<$0.01} \\
\end{tabular}
\end{table}

\begin{table}[h]
\begin{center}
\caption{Two robustness checks for the regression discontinuity design. In the first column, we create a fake arbitrary threshold at 6 million NOK, with firms just under the threshold labeled as treated, while those just over are labeled as the control. In the second column, we identify the firms that were just under the threshold in 2012 and identify them as treatment, but use data from 2010, before the implimentation of the policy. Both regressions limit the data to firms that are within 1 million NOK of the working capital threshold and include fixed effects for sector. In both cases, the treatment variable is estimated to be close to zero and insignificant. }
\begin{tabular}{@{\extracolsep{5pt}}lcc}
\\[-1.8ex]\hline
\hline \\[-1.8ex]
 & \multicolumn{2}{c}{\textit{Dependent variable:}} \\
\cline{2-3}
\\[-1.8ex] & \multicolumn{2}{c}{Dividend} \\
\\[-1.8ex] & I & II\\
\hline \\[-1.8ex]
Intercept          &  -0.448*** &    -0.416 \\
                   &    (0.141) &   (0.971) \\
noAudit           &     -0.036 	 &   -0.017 \\
                   &    (0.058) &   (0.084) \\
operating revenue    &      0.204 &     0.095 \\
                   &    (0.185) &   (0.204) \\
employees     &  -0.083*** &    -0.016 \\
                   &    (0.018) &   (0.066) \\
total assets         &   0.170*** &     0.059 \\
                   &    (0.034) &   (0.094) \\
risk (roa)           &   -0.037** &    -0.014 \\
                   &    (0.018) &   (0.052) \\
leverage           &   0.146*** &     0.090 \\
                   &    (0.026) &   (0.057) \\
cash flow (mu)     &   0.4567***  &  0.286*** \\
                   &     (0.029)  &   (0.073) \\
cash flow (sd)     &   0.0393*    &    -0.015 \\
                   &      (0.023)  &   (0.064) \\
\hline
Sector FE 			&      Yes &        YES \\
R2                 &       0.12 &      0.28 \\
Adjusted R2        &       0.10 &      0.18 \\
N                  &       2861 &       398 \\
RMSE               &       0.83 &      0.59 \\
\bottomrule
\end{tabular}
\end{center}
\end{table}

\FloatBarrier

\section*{Appendix C: Difference-in-difference}\label{Appendix C}

The regression discontinuity design takes advantage of the arbitrariness of the policy thresholds on firm size. However, the design does not make full use of the panel of data to inform the estimates, as only data after the introduction of the policy is used.

In this section, we make use of data prior to the introduction of the policy in order to aid in estimating an unbiased treatment effect. The basic intuition is that by comparing not just between treatment and control groups but also before and after the introduction of the policy, the effect of unobserved time-invariant covariates that affects selection into the treatment group can be implicitly controlled for.

Following the notation of \citet{angrist_mostly_2009}, consider the simple model:

%Firm i, state s,
\begin{equation}
DIV_{ist} = \gamma_s + \lambda_t + \tau NoAudit_{st} + \epsilon_{ist}
\label{eq:diffindiff1}
\end{equation}

Here $DIV_{ist}$ represents the dividend for a firm i, in state s (receiving or not receiving an audit), at time t (2012, the first year with a full roll-out of the policy or 2010, the last year of the old regime).\footnote{As can be seen visually in figure\ref{fig:scatter_size}, 2011 had a limited role-out of the policy} $\gamma_s$ represents the state effect on dividend payout, while $\lambda_t$ represents the time period effect. $noAudit_{st}$ is the indicator variable for the policy, where $\tau$ again represents the effect of the policy.

Taking the expectation and differencing across periods for both treatment and control groups gives:

\begin{align}
E[DIV_{ist}|s=Audit, t=2012] - E[Y_{ist}|s=Audit, t=2010] = \lambda_{2012} -\lambda_{2010} \\
E[DIV_{ist}|s=NoAudit, t=2012] - E[Y_{ist}|s=NoAudit, t=2010] = \lambda_{2012} -\lambda_{2010} + \tau
\end{align}

%population difference-in-difference
The population difference-in-difference estimator can then be written:

\begin{align}
\begin{split}
{E[DIV_{ist}|s=Audit, t=2012] - E[Y_{ist}|s=Audit, t=2010]} &- \\
{E[DIV_{ist}|s=NoAudit, t=2012] - E[Y_{ist}|s=NoAudit, t=2010]} &= \tau
\end{split}
\end{align}

\subsection{Results, difference-in-difference}

We seek to improve the efficiency of the difference-in-difference estimator by controlling for continuous covariates that vary across the threshold. This includes the number of employees, operating revenue, total assets and a measure of risk (return on assets).

The regression equation can be written:

\begin{equation}
\Delta DIV_{is} = \alpha + \tau noAudit_{s} + \mathbf{SIZE \zeta} + \mathbf{X\beta} + \epsilon_{is}
\end{equation}

Here $\Delta DIV_{is}$ represents the first difference for firm dividends between years 2012 and 2010. we drop the t index, as the relevant treatment indicator and covariates are for 2012, the year when the policy was fully put in place. The treatment indicator can then be interpreted as the change in firm dividend between 2012 and 2010 in the treatment group, relative to the change in dividend in the control group and conditional on the observable covariates represented by the matrices $\mathbf{SIZE}$ and $\mathbf{X}$.

Table \ref{tbl:diff-in-diff} shows results from the estimation. The first column shows results where only an intercept and the treatment effect are included. In the second column, covariates of firm size, risk and solidity are included. Mean and standard deviations of cash flow are also included. The point estimate moves from being negative in the first column to positive and significant at the 10\% level in the second column. In the third and fourth columns, the data is limited to firms with operating revenue between 4 and 6 million NOK, where the fourth column also includes sector fixed effects. Limiting the sample to firms between 4 and 6 million leads to a higher point estimate of the treatment effect, though the inclusion of fixed effects reduces the point estimate and increases the standard error.

\begin{table}[ht]

\caption{Difference-in-Difference estimators. The left-hand-side variable is the difference in dividend between 2010 (pre-reform) and 2012 (post-reform). The first two columns show regressions using the full dataset. The second column includes control variables. The third and fourth columns show results from a regression with only firms with operating revenue between 4 and 6 million. The fourth column shows results from a model that includes sector fixed effects.\label{tbl:diff-in-diff}}
\begin{center}
\begin{tabular}{lccccc}
\hline
                   & I & II & III & IV  \\
\hline
Intercept           & 0.017   & -0.082** & -0.170*** & -0.314     \\
                    & (0.023) & (0.041)  &  (0.052)   & (0.736)    \\
noAudit             & -0.095* & 0.125*   & 0.201*    & 0.141      \\
                    & (0.055) & (0.074)  &  (0.109)   & (0.119)    \\
operating revenue     &         & 0.072**  &  0.508**   & 0.418      \\
                    &         & (0.031)  &  (0.243)   & (0.262)    \\
employees     &         & 0.006    &  -0.061    & -0.108*    \\
                    &         & (0.023)  &  (0.049)   & (0.064)    \\
total assets           &         & 0.018    &  0.157***  & 0.168***   \\
                    &         & (0.026)  &  (0.048)   & (0.051)    \\
risk\_roa           &         & -0.028   & 0.104**   & 0.121**    \\
                    &         & (0.027)  & (0.048)   & (0.052)    \\
cash\_flow\_mu      &         & 0.152*** &0.139**   & 0.186***   \\
                    &         & (0.023)  &(0.060)   & (0.070)    \\
cash\_flow\_sd      &         & 0.008    &  -0.239*** & -0.324***  \\
                    &         & (0.029)  & (0.075)   & (0.086)    \\

\hline
Sector FE 			& NO & NO &  NO & YES \\
N                   & 2231    & 2231     & 374       & 374        \\
R2                  & 0.00    & 0.03     & 0.06      & 0.13       \\
Adjusted R2         & 0.00    & 0.03     & 0.05      & -0.00      \\
\hline
\end{tabular}
\end{center}
\end{table}

The results from the difference-in-difference model were estimated with more uncertainty, which is to be expected given that the left-hand-side variable is now a differenced variable, and we only use data from two years. Nonetheless, the sign and magnitude of the point-estimates are in line with the results from the regression discontinuity design and adds to the evidence that allowing firms to forgo an auditor leads to on average higher dividend pay-out.

\end{document}